\documentclass[11pt,a4paper]{article}
\pdfoutput=1
\usepackage{pub}
\usepackage[T1]{fontenc}
\usepackage[english]{babel}
\usepackage{lmodern}
\usepackage{bbm}
\usepackage{bm}
\usepackage{color}
\usepackage{slashed}
\usepackage{graphicx}
\usepackage{braket}
\usepackage{bbold}
\usepackage{dsfont}

\DeclareGraphicsExtensions{.pdf,.png,.jpg}
\DeclareMathOperator{\arccot}{arccot}

\allowdisplaybreaks[1]
\def\simg{{\ \lower-1.2pt\vbox{\hbox{\rlap{$>$}\lower6pt\vbox{\hbox{$\sim$}}}}\ }}
\def\siml{{\ \lower-1.2pt\vbox{\hbox{\rlap{$<$}\lower6pt\vbox{\hbox{$\sim$}}}}\ }}

\makeatletter \@addtoreset{equation}{section} \makeatother

\newcommand{\vrel}{v_{\text{rel}}}
\newcommand{\mD}{m_{\text{D}}}
\newcommand{\Tref}{T_{\textrm{ref}}}

\begin{document}

\flushbottom

\begin{titlepage}

	\begin{centering}
		
		\vfill

		{\Large{\bf
				Interplay between improved interaction rates and modified cosmological histories for dark matter
		}} 
		
		\vspace{0.8cm}
		
		S.~Biondini

		\vspace{0.8cm}
		
		{\em Department of Physics, University of Basel,
			\\
			Klingelbergstr. 82, CH-4056 Basel, Switzerland} 
		
		\vspace*{0.8cm}

	\end{centering}
	
	\vspace*{0.3cm}
	
	\noindent
	
	\textbf{Abstract}:  A novel particle has been and still is an intriguing option to explain the strong evidence for dark matter in our universe. To quantitatively predict the dark matter energy density, two main ingredients are needed: interaction rates and an expansion history of the universe. In this work, we explore the interplay between recent progress in the determination of particle production rates and modified cosmological histories. For the freeze-out mechanism, we focus on Sommerfeld and bound-state effects, which boost and make dark matter pair annihilation more efficient. As regards the freeze-in option, we include thermal masses, which enter the decay processes that produce dark matter, and we find that they can suppress or enhance the dark matter yield. We consider a class of modified cosmological histories that induce a faster universe expansion, and we assess their effect in combination with improved particle interaction rates on the dark matter energy density.

	\vfill
	
\end{titlepage}

\thispagestyle{empty}
\setcounter{page}{0}
\tableofcontents
\clearpage

\section{Introduction}

One of the major open challenges across cosmology and particle physics is to understand the content of our universe.
There is mounting evidence that the fundamental building blocks of the Standard Model of Particle Physics (SM) account only for a small fraction of the matter in the cosmos,
whereas the bulk appears to be some sort of non-luminous and non-baryonic particles.
Complementary measurements of galaxy formation, gravitational lensing,  large scale structures,  and the cosmic
microwave background (CMB) point to the astounding conclusion that more than 80\% of the matter consists of dark matter (DM). Nowadays, the DM energy density is  very accurately determined by temperature anisotropies
of the CMB and it amounts to $\Omega_{\textrm{DM}} h^2 = 0.1200 \pm 0.0012$~\citep{Planck:2018nkj}, where $h$ is the reduced Hubble constant.

Although this is not the only viable option, it is well possible for DM to be a new, yet undiscovered, particle. The fervent interplay with particle-physics driven motivations for new physics has produced a plethora of models and rather different DM candidates,  
see e.g.~\cite{Bertone:2004pz,Feng:2010gw} for extensive reviews. 
In order to establish whether a given DM model is cosmologically viable, one has to compare the corresponding prediction for the energy density and check its consistency with the Planck measurement. In practice one needs to link the Lagrangian field content and parameters, most notably masses and couplings, with a \textit{production mechanism} in the early universe. 
In this work, we shall consider the \textit{freeze-out} \cite{Lee:1977ua,Griest:1990kh,Gondolo:1990dk} and \textit{freeze-in} \cite{Moroi:1993mb,McDonald:2001vt,Hall:2009bx} mechanisms. 
In the former scenario, dark matter particles follow an equilibrium abundance when the
temperature is larger than their mass and are kept in chemical equilibrium via pair annihilation, which are very efficient up until $T/M \approx 1/25$. Around this temperature, dark matter particles decouple from the thermal bath and their abundance is frozen ever since. The freeze-in mechanism entails the opposite situation: dark matter particles never reach equilibrium due to very
small couplings with the plasma constituents. Typically dark matter particles are generated through the decays of heavier accompanying states in the dark sector, as well as $2 \to 1$ annihilations and $2 \to 2$
scatterings that may involve SM particles. Dark matter particles only appear in the final
state of the relevant processes, and its abundance increases over the thermal
history. For renormalizable operators, the more important temperature window for freeze-in production is $T \gtrsim M$, which is complementary to that typical of the freeze-out scenario. 

Both for the freeze-out and  freeze-in mechanisms, the interaction rates that are used for the prediction of the DM energy density are sometimes incomplete even at leading order. This is mainly due to a rather involved field content of realistic, and non-minimal, dark matter models, which triggers a series of compelling (thermal) phenomena. In this work, we want to consider two exemplary situations:  (i) non-perturbative effects on non-relativistic annihilations for the freeze-out mechanism and (ii) the role of thermal masses in $1 \to 2$ decays driving the freeze-in production.

Around the freeze-out temperature, dark matter particles are non-relativistic and slowly moving objects in the early universe thermal environment. Such a kinematical condition calls for a scrutiny and the inclusion of \textit{near-threshold} (or \textit{non-perturbative}) effects, that may be rather impactful  depending on the details of the particle physics model. In many models, DM particles or accompanying states of the dark sector often interact with gauge bosons or scalar particles, which trigger self interactions between dark sector particles. Typical manifestations of long-range interactions as induced by a repeated mediator exchange in the \emph{soft}-momentum region are the Sommerfeld enhancement (for an attractive potential) \cite{Sommerfeld,Hisano:2004ds}  and bound-state formation (BSF) \cite{Detmold:2014qqa,vonHarling:2014kha}. The latter is due to transitions of DM pairs from a scattering state (or above-threshold state) into a bound state (or below-threshold state), and it can occur via different processes in a thermal environment \cite{vonHarling:2014kha,Kim:2016kxt,Biondini:2018pwp,Binder:2019erp}. 
In the non-relativistic regime, Sommerfeld factors and bound-state formation are  formally a leading order effect. Unless the coupling between the DM particles and force carriers is quite small,
 the inclusion of Sommerfeld and bound-state formation is crucial for a correct estimation of the DM energy density. Especially bound-state formation and decays work as an additional efficient channel to deplete the DM population. The effective annihilation cross section is increased and one typically finds larger DM masses that are compatible with the measurement of the Planck satellite for a fixed value of the DM couplings.
Recent and ongoing efforts have shown that bound-state effects can substantially change the model parameter space that is compatible with the observed energy density. A research program that aims to reassess the reach of present and forthcoming experiments searching for DM, namely direct and indirect detection, as well as collider searches has started only very recently, see e.g.~refs. \cite{Laha:2015yoa,Asadi:2016ybp,Garny:2018ali,Biondini:2018ovz,Biondini:2019int,Garny:2021qsr,Bottaro:2021srh,Becker:2022iso,Biondini:2023ksj} for exemplary studies. 

As far as the less explored freeze-in mechanism is concerned, a systematic derivation of thermal cross sections and widths is also a topic of ongoing research. Here, the relativistic ($T \sim M$) and ultra-relativistic ($T \gg M$) regimes are relevant, hence anticipating a prominent impact from plasma effects. Only recently the use of a Maxwell-Boltzmann distribution has been replaced by a more appropriate Fermi-Dirac/Bose-Einstein distribution for  the decaying particle~\cite{Belanger:2018ccd,Lebedev:2019ton,Bandyopadhyay:2020ufc}.  Despite the actual DM particle is feebly interacting, it is usually produced in  multi-particle collisions or decays  of equilibrated states in the thermal environment. This condition, which is largely model-independent, calls for a scrutiny of various thermal effects that are triggered by the interactions responsible for the equilibrium of such states, either SM gauge interactions or those of some hidden sector. Most notably, frequent interactions with a dense medium induce thermal masses and multiple soft scatterings. The latter, which is  oftentimes called the Landau--Pomeranchuk--Migdal (LPM) effect,  typically enhances the $1 \to 2$ decays, and makes other effective $1 \leftrightarrow 2$ processes possible \cite{Anisimov:2010gy,Besak:2012qm,Ghisoiu:2014mha}. At high temperature, $2 \to 2$ scatterings have to be treated with care when a soft-momentum region appears in the relevant processes \cite{Besak:2012qm,Ghiglieri:2016xye}. For all the mentioned processes, thermal masses enter as a key ingredient and this is the aspect we focus on in this work.

A thermal interaction rate  on its own is almost meaningless unless it is compared with the expansion rate of the universe, namely the Hubble rate. Irrespective of the production mechanism, either departing from thermal equilibrium (freeze-out) or never reaching it (freeze-in), the Hubble rate sets the clock that measures the efficiency of particle interactions. 
The main difficulty is that we do not know much about the expansion history of the universe at epochs prior to the Big Bang Nucleosynthesis (BBN). 
Even though the early universe had to be radiation dominated at the onset of BBN, which occurs at $\mathcal{O}(1)$ MeV temperatures, any scenario that implies higher temperatures should admit the possibility for \emph{different} cosmological histories. The common lore is to extrapolate the condition of the early universe at the BBN backward in time and at (much) larger temperatures. As we usually are open-minded about the diversity of new-physics models, we owe the early universe the same. 
The implications of a modified expansion rate has been considered in the literature for many DM models and the impact on the DM yield can be quite large, see e.g.~\cite{Tenkanen:2016jic,Hardy:2018bph,Bernal:2018kcw,Bernal:2018ins,Arias:2020qty} for scalar and fermionic singlet dark matter, Higgs and $Z$ portal models \cite{Chanda:2019xyl,Bernal:2019mhf,Chang:2021ose}, inert doublet and triplet scalar dark matter models \cite{Barman:2021ifu},  Higgsino \cite{Han:2019vxi} and  neutralino dark matter \cite{Salati:2002md,Profumo:2003hq,Drees:2018dsj},  asymmetric dark matter \cite{Sujuan:2023lne} and axion-like particles \cite{Ghosh:2023tyz}.
Here we aim to explore the interplay between ameliorated thermal rates with modified cosmological histories.  This is the main original contribution of the present work. More specifically, we include near-threshold effects for the thermal freeze-out and thermal masses for freeze-in produced dark matter, which may enhance or reduce the corresponding thermal rates, and combine them with a modified expansion history. 

The paper is organised as follows. In section~\ref{sec:NT_effects}, we describe near-threshold effects  for DM pairs within the framework of potential non-relativistic effective field theories (pNREFTs). We discuss two exemplary models with a vector and scalar force carrier respectively, and highlight similarities and differences. Freeze-in via $1\to 2$ decays is addressed in section~\ref{sec:thermal_masses_FI} for $t$-channel DM models, where we include thermal-mass effects in the DM production rate. Section~\ref{sec:mod_cosmo} is devoted to a self-contained summary of modified cosmological histories that feature a faster expansion rate of the universe before the BBN epoch. Numerical results that show the interplay between improved thermal rates and modified cosmological histories are discussed in section~\ref{sec:numerics}, whereas conclusions and outlook are offered in section~\ref{sec:conclusions}. 

\section{Near-threshold effects for DM freeze-out}
\label{sec:NT_effects}
Non-relativistic DM particles are susceptible to a non-trivial dynamics whenever they interact through some force carrier. This happens in many ultraviolet completions of the SM, as well as in simplified dark matter models, where DM particles and/or coannihilating partners interact via gauge bosons or scalar fields.  If the mediator is sufficiently light to induce long-range interactions, DM pair annihilations can be severely affected. Typically, the Sommerfeld enhancement increases the annihilation cross section for
a pair in an attractive channel, that implies larger dark matter masses as compatible with
the observed relic density. Moreover, there is yet another manifestation of multiple soft exchanges, namely, the existence of bound states: whenever they form in the early universe, and they are not efficiently dissociated, DM  can be depleted also via bound-state decays. Hence, an additional efficient annihilation channel is active and, for a fixed value of the coupling strength, the corresponding dark matter energy density gets further reduced.

In order to illustrate such effects, we consider the two following models: (i) Dirac dark matter fermion with a vector mediator in section~\ref{sec:abelian_model}; (ii) Dirac dark matter fermion with a scalar mediator in section~\ref{sec:scalar_med_model}. We first discuss some general features of non-relativistic dark matter pairs in a thermal environment within the framework of non-relativistic effective field theories (NREFTs) and potential NREFTs (pNREFTs). Then, we specify the form of the low-energy theories for the two models and list the main observables that are necessary for the determination of the DM energy density.   

\subsection{pNREFTs for dark matter, Sommerfeld factors and bound-state formation}
\label{sec:somm_and_bsf}

The treatment of interacting non-relativistic particle pairs in a thermal environment is rather complicated because of the presence of many energy scales. 
To begin with, there are the dynamically generated scales by the relative motion: (i)
the momentum transfer, which is also proportional to the inverse of the typical size of the pair; (ii) the kinetic/binding energy of the pair. Such scales are hierarchically ordered with the DM mass  for 
near threshold states moving with relative non-relativistic velocities $v_{\hbox{\scriptsize rel} }$, namely $M \gg M v_{\hbox{\scriptsize rel} } \gg Mv_{\hbox{\scriptsize rel} }^2$.
The relative velocity of the pair is fixed by the virial theorem to be $v_{\hbox{\scriptsize rel} } \sim \alpha$ for Coulombic bound states.
Therefore, the corresponding hierarchy is $M \gg M \alpha \gg M \alpha^2$, 
where $\alpha = g^2 / (4 \pi)$ is the fine structure constant in terms of the coupling $g$ between the DM particle and the force mediator. The in-vacuum scales are useful to define respectively the hard, soft and ultrasoft energy modes of a given particle theory. 
Along with the in-vacuum scales, there are thermodynamical scales, namely the plasma temperature $T$ and the Debye mass $\mD$, which is the inverse of the chromoelectric screening length; for a weakly coupled plasma $\mD \sim gT$. We do not include the effect of thermal masses in the following treatment of the freeze-out. 
Such a multi-scale system is well suited for a treatment in terms of Effective Field Theories (EFTs). We assume the following hierarchy of scales 
\begin{eqnarray}
    M \gg M \alpha \gg M \alpha^2 \gtrsim T \, ,
    \label{H_scales}
\end{eqnarray}
that we write specifically for the bound states.\footnote{Depending on the details and degrees of freedom of the DM model, $T \gtrsim M \alpha^2$ may induced thermal masses and modify the Coulomb potential, see e.g.~\cite{Kim:2016kxt,Biondini:2018pwp,Binder:2018znk} and \cite{Laine:2006ns,Brambilla:2008cx} for former studies about heavy quarkonium. }

In this work, we exploit the framework of NREFTs~\cite{Caswell:1985ui,Bodwin:1994jh} and pNREFTs~\cite{Pineda:1997bj,Brambilla:1999xf} when dealing with interacting dark matter pairs and the observables of interest, namely cross sections and widths. We find convenient indicating pairs in a scattering state with $(X \bar{X})_p$, where $p=M \vrel/2$ denotes the momentum of the relative motion, whereas a fermion-antifermion pair in a bound state is indicated with $(X \bar{X})_n$. In order to shorten the notation, $n \equiv|n\ell m \rangle$ stands for the set of quantum number of a given bound state. The main relevant processes include DM pair annihilations into light mediators (scalar or vector fields), which can occur both for scattering states $(X \bar{X})_p$ and bound states $(X \bar{X})_n$, and bound-state formation. For a detailed derivation and discussion of the pNREFTs for vector and scalar mediators, with an explicit application to dark matter, see refs.~\cite{Beneke:2014gja,Binder:2020efn,Biondini:2021ycj,Biondini:2023zcz} and references therein. We summarize here the main steps and streamline the derivation of the towers of low-energy theories. 

Since the ultimate goal is to address near-threshold effects at the ultrasoft scale, we shall integrate out the hard and soft energy modes in a two-step construction of the low-energy Lagrangian. 
The first step accounts for integrating out the hard energy/momentum modes of order $M$. The corresponding low-energy theory, which we generically indicate with NREFT$_{\textrm{DM}}$, describes non-relativistic dark
fermions and antifermions and low energy mediators, and it is organized as an expansion in $1/M$ and $\alpha$. The Lagrangian splits into a bilinear and a four-fermion sectors. The former comprises interactions between non-relativistic fermion (and antifermions) with the force mediator. The four-fermion Lagrangian is especially relevant because the imaginary part of the corresponding matching coefficients originate from the
particle-antiparticle annihilation diagrams \cite{Bodwin:1994jh}. In the so-obtained EFT the soft scale ($M\alpha$) and ultrasoft scales ($M \alpha^2$ and the temperature) are still intertwined.

The next step is to integrate out the typical relative distance among fermions and antifermions, which is induced by the soft-momentum exchange of the force mediator. As a result,  the degrees of freedom are \textit{interacting} dark matter pairs and ultrasoft mediators (see figure~\ref{fig:FrPr_fig1.pdf} for a diagrammatic representation). The so-obtained low-energy theory is dubbed as pNREFT$_{\textrm{DM}}$. The potential between a fermion and an antifermion appears as a matching coefficient and it is derived in a field theoretical fashion, namely relativistic and quantum corrections can be computed. Moreover, there is a power counting that helps in estimating contributions to a given observable.  
Threshold phenomena affect fermion-antifermion pairs, hence it is convenient to project the EFT on the fermion-antifermion space
and express it in terms of a fermion-antifermion bilocal field $\varphi(t,\bm{r},\bm{R})$,\footnote{In order to clarify on the distinction between soft and ultrasoft mediators, and to introduce the degrees of freedom of pNREFT$_{\textrm{DM}}$, we project onto the particle-antiparticle sector as follows, $
\int d^3 \bm{x}_1 d^3 \bm{x}_2 \varphi_{ij}(t,\bm{x}_1, \bm{x}_2) \psi^\dagger_i (t,\bm{x}_1) \chi_{c,j}^\dagger (t,\bm{x}_2) | \Phi_{\textrm{US}}\rangle $, 
where $i,j$ are spin indices, while the state $| \Phi_{\textrm{US}}\rangle$ contains no heavy particles/antiparticles and an arbitrary number of mediators with energies much smaller than $M\alpha$.}
where $\bm{r} \equiv \bm{x}_1-\bm{x}_2$ is the distance between a fermion at $\bm{x}_1$ and an antifermion at $\bm{x}_2$, which is typically of order $1/(M\alpha)$, 
and $\bm{R}\equiv(\bm{x}_1+\bm{x}_2)/2$ is the center of mass coordinate, which is of order $1/(M\alpha^2)$.
In order to ensure that the mediators are ultrasoft, the corresponding fields are \textit{multipole expanded} in $\bm{r}$.
Hence, a generic pNREFT$_\textrm{DM}$ Lagrangian density is organized as an expansion in $1/M$ and $\alpha(M)$,
inherited from NREFT$_\textrm{DM}$, and an expansion in $\bm{r}$ and $\alpha(1/r)$ and it reads schematically
\begin{eqnarray}
  \mathcal{L}_{\textrm{pNREFT}_{\textrm{DM}}}&=&   \int d^3 \bm{r} \; \varphi^\dagger(t,\bm{r},\bm{R})
             \, \left[ i \partial_0 -H(\bm{r},\bm{p},\bm{P},\bm{S}_1,\bm{S}_2)  \right] \varphi (t,\bm{r},\bm{R})   \nonumber \\
&&+ \mathcal{L}^{\textrm{int}}_{\textrm{ultrasoft}}(\varphi(t,\bm{r},\bm{R}),\Phi(t,\bm{R})) + \mathcal{L}^{\textrm{mediator}}_{\textrm{ultrasoft}}(\Phi(t,\bm{R}))
 \, ,
\label{pNREFT_generic}
\end{eqnarray}
where 
\begin{eqnarray}
 &&H(\bm{r},\bm{p},\bm{P},\bm{S}_1,\bm{S}_2) =  \frac{\bm{p}^2}{M}+\frac{\bm{P}^2}{4M} - \frac{\bm{p}^4}{4M^3} +  V (\bm{r},\bm{p},\bm{P},\bm{S}_1,\bm{S}_2) + \ldots\, , 
 \label{ham_pNREFT_DM}\\
  &&V (\bm{r},\bm{p},\bm{P},\bm{S}_1,\bm{S}_2)= V^{(0)} + \frac{V^{(1)}}{M} + \frac{V^{(2)}}{M^2} + \frac{V^{(4)}}{M^4} \ldots \, ,
 \label{pot_pNREFT_DM}    
\end{eqnarray}
and $\bm{S}_1=\bm{\sigma}_1/2$ and $\bm{S}_2=\bm{\sigma}_2/2$ are the spin operators acting on the fermion and antifermion, respectively. Then in the second line of eq.~\eqref{pNREFT_generic} we display (i) the interaction Lagrangian that involves the bilocal field and the mediator $\Phi(t,\bm{R})$, the latter does not depend on $\bm{r}$; (ii) the Lagrangian term that comprises only mediator fields, which have been multiple expanded. 

In the limit of massless mediators that we consider in this study, the leading order term $V^{(0)}$ in eq.~\eqref{pot_pNREFT_DM} is the Coulomb potential.  The imaginary part of the potential terms $V^{(2)}/M^2$ and $V^{(4)}/M^2$ consists of local operators and describe the annihilations process $(X \bar{X})_p \to \Phi \Phi$ for a scattering state as well as $(X \bar{X})_n \to \Phi \Phi$ for a given bound state.
 We single out from $V (\bm{r},\bm{p},\bm{P},\bm{S}_1,\bm{S}_2)$ in eq.~\eqref{pot_pNREFT_DM} the annihilation terms up to order $1/M^4$, which account for the annihilation of scattering and bound states in $S$- and $P$ waves~\cite{Brambilla:2002nu,Brambilla:2004jw} 
 \begin{eqnarray}
\mathcal{L}^{\textrm{ann}}_{\hbox{\tiny pNREFT}_{\textrm{DM}}}&=& \frac{i}{M^2} \, \int d^3 \bm{r} \varphi^\dagger (\bm{r}) \delta^3(\bm{r}) \left[ 2 {\rm{Im}}[f(^1S_0)] - \bm{S}^2 \left( {\rm{Im}}[f(^1S_0)]-  {\rm{Im}}[f(^3S_1)] \right) \right] \varphi (\bm{r}) 
\nonumber
\\
&&+\frac{i}{M^4} \, \int d^3 \bm{r} \varphi^\dagger (\bm{r}) \mathcal{T}_{SJ}^{ij} \nabla_{\bm{r}}^i \delta^3(\bm{r})  \nabla_{\bm{r}}^j \,  {\rm{Im}}  [f(^{2 S+1}P_{J})] \varphi \, (\bm{r}) 
\nonumber
\\
&&+\frac{i}{2M^4} \, \int d^3 \bm{r} \varphi^\dagger (\bm{r}) \,  \Omega_{SJ}^{ij} \left\lbrace \delta^3(\bm{r}),\nabla_{\bm{r}}^i \nabla_{\bm{r}}^j  \right\rbrace  {\rm{Im}} [g(^{2 S+1}S_{J})] \varphi \, (\bm{r}) \, ,
\label{pNREFT_ann_generic}
 \end{eqnarray}
 where $\bm{S}$ is the spin of the pair ($\bm{S}^2=0$ for spin singlets and $\bm{S}^2=2$ for spin triplets), while $\mathcal{T}_{SJ}^{ij}$ and $\Omega_{SJ}^{ij}$ are spin projector operators (cfr.~e.g.~\cite{Brambilla:2002nu,Brambilla:2004jw}). We did not write the $\bm{R}$ and $t$ dependence in the argument of the field $\varphi$ to avoid cluttering the notation.
  The spectroscopic notation is $^{2S+1} L_J$ where $S$, $L$ and $J$ are respectively the spin, orbital angular momentum and total momentum of the annihilating pair. It turns out to be quite useful to identify each partial-wave contribution to the pair annihilations, so that one can easily associate Sommerfeld factors for the scattering states. By computing the annihilation cross section for the scattering states and the decay width for the bound states in pNREFT$_{\textrm{DM}}$, the factorization between hard and soft modes is made manifest. Multiple Coulomb scatterings are encoded in the wave function of the annihilating pair. 

The bound state formation process $(X \bar{X})_p \to (X \bar{X})_n + \Phi$ is  triggered by the ultrasoft interaction $\mathcal{L}^{\textrm{int}}_{\textrm{ultrasoft}}(\varphi(t,\bm{r},\bm{R}),\Phi(t,\bm{R}))$, and its explicit form depends on the relativistic theory, or equivalently, on the dark matter model one starts with. In this paper, we focus on the bound-state formation via the radiative emission of the mediators (for complementary bound-state formation processes, which demand a richer dark sector, see \cite{Kim:2016kxt,Biondini:2018pwp,Binder:2019erp,Binder:2020efn}). Ultrasoft interactions are also responsible for transitions among different bound states, that may further boost the relevance of bound-state effects for the DM energy density \cite{Garny:2021qsr,Biondini:2021ycj,Binder:2021otw,Biondini:2023zcz,Binder:2023ckj}.

\subsection{Dark matter with a vector mediator}
\label{sec:abelian_model}
In this section, we consider a simple model where the dark sector consists of a dark Dirac fermion $X$ that is charged under an abelian gauge group~\cite{Feldman:2006wd,Fayet:2007ua,Goodsell:2009xc,Morrissey:2009ur,Andreas:2011in}.
The Lagrangian density reads
\begin{equation}
\mathcal{L}=\bar{X} (i \slashed {D} -M) X -\frac{1}{4} F_{\mu \nu} F^{\mu \nu} + \mathcal{L}_{\textrm{portal}} \, ,
\label{lag_mod_0_vec}
\end{equation}
where the covariant derivative is $D_\mu=\partial_\mu + i g V_\mu$, $V_\mu$ is the vector field and $F_{\mu \nu} = \partial_\mu V_\nu - \partial_\nu V_\mu$;
we define the corresponding fine structure constant $\alpha \equiv g^2/(4 \pi)$.
The term $\mathcal{L}_{\textrm{portal}}$ encompasses additional interactions of the dark photon with the SM degrees of freedom.
A common realisation of the portal interaction is a mixing with the neutral components of the SM gauge fields \cite{Holdom:1985ag,Foot:1991kb}.
Such interactions are responsible for the eventual decay of the dark photons,
so that their number density does not dominate the universe at later stages. As far as this work is concerned, we do not consider the portal interaction term and neglect it in the following. 

The low-energy theory at the ultrasoft scale, which is obtained from the model Lagrangian \eqref{lag_mod_0_vec}, comes in the form of pNRQED~\cite{Pineda:1997bj,Pineda:1997ie}. The scrutiny of the corresponding derivation in the context of dark matter can be found in ref.~\cite{Biondini:2023zcz}. By integrating out the hard and soft scales, one arrives at the following Lagrangian
\begin{eqnarray}
\mathcal{L}_{\textrm{pNRQED}_{\textrm{DM}}}&=&   \int d^3r \; \varphi^\dagger(t,\bm{r},\bm{R})
             \, \left[ i \partial_0 -H(\bm{r},\bm{p},\bm{P},\bm{S}_1,\bm{S}_2)  + g \, \bm{r} \cdot \bm{E}(t,\bm{R})\right] \varphi (t,\bm{r},\bm{R})  
             \nonumber
             \\&-&\frac{1}{4} F_{\mu \nu} F^{\mu \nu} \, ,
\label{pNREFT_vector_tot}
\end{eqnarray}
where we added the subscript in order to remind that the low-energy theory is for the abelian dark matter model in eq.~\eqref{lag_mod_0_vec}, and not for QED. 
We notice that $\mathcal{L}_{\textrm{pNRQED}_{\textrm{DM}}}$ is of the general form as in eqs.~\eqref{pNREFT_generic} and \eqref{pNREFT_ann_generic}. Dark matter pair annihilation is accounted for in the imaginary part of the local potential in $H(\bm{r},\bm{p},\bm{P},\bm{S}_1,\bm{S}_2)$ and reorganised in eq.~\eqref{pNREFT_ann_generic}. The matching coefficients of the four-fermion operators of NRQED read, at order $\mathcal{O}(\alpha^2)$, as follows~\cite{Bodwin:1994jh,Vairo:2003gh}
\begin{eqnarray} 
       {\rm{Im}}[f(^1S_0)] = \pi \alpha^2 \, , \quad  {\rm{Im}}[g(^1S_0)] = -\frac{4}{3} \pi \alpha^2 \, ,
       \\
        {\rm{Im}}[f(^3P_0)] = 3 \pi \alpha^2 \, , \quad  {\rm{Im}}[f(^3P_2)] = \frac{4}{5} \pi \alpha^2 \, .
\end{eqnarray}
We only display the non-vanishing matching coefficients. 
Depending on the two-particle states one projects onto, respectively scattering and bound states, one obtains a cross section or a decay width. The annihilation cross section manifestly shows the factorization of hard and soft contributions and it reads
  \begin{align}
    & \sigma_{\textrm{ann}} v_{\textrm{rel}}((X\bar{X})_p \to \gamma \gamma)  =
     \left( \frac{{\rm{Im}}[f(^1S_0)]}{M^2} + \frac{p^2 \, {\rm{Im}}[g(^1S_0)]}{M^4} \right) |\mathcal{R}_{0}(0)|^2 
     \nonumber
     \\
     &~~~+ \frac{{\rm{Im}}[f(^3P_0)]+5{\rm{Im}}[f(^3P_2)]}{3 M^4}   \big \vert \mathcal{R}'_{1}(0) \big \vert^2 
 = \frac{ \pi \alpha^2}{M^2} \left( 1 - \frac{v^2_{\textrm{rel}}}{3}\right) S_0(\zeta)  + \frac{7 \pi  \alpha^2 v_{\textrm{rel}}^2}{12 M^2} S_1(\zeta) \, ,    
     \label{ann_cross_Sommerfeld_vec}
 \end{align}
 where $\mathcal{R_{\ell}}(r)$ is the radial wave function of a Coulombic scattering state with $\ell=0,1$, and the $S$- and $P$-wave Sommerfeld factors are connected to the squared wave function via
 \begin{eqnarray}
 \big \vert \mathcal{R}_{0}(0) \big \vert^2=\frac{2 \pi \zeta}{1-e^{-2 \pi \zeta}} \equiv S_0(\zeta) \, , \quad  \big \vert \mathcal{R}'_{1}(0) \big \vert^2 =p^2 S_0(\zeta) (1+\zeta^2) \equiv p^2 S_1(\zeta) \, ,
 \label{som_factors}
 \end{eqnarray}
where $\zeta=\alpha/\vrel$. 
The corresponding observable for a bound state is a decay width.  The expressions for $n$S and $n$P states exhibit the analogous hard versus soft factorization 
\begin{eqnarray}
     \Gamma_{\textrm{ann}}^{nS}((X\bar{X})_n \to \gamma \gamma)  =\frac{|R_{nS}(0)|^2}{\pi M^2} \left\lbrace   {\rm{Im}}[f(^1S_0)] + \frac{E_{n}}{M}   {\rm{Im}}[g(^1S_0)]  \right\rbrace = \frac{M  \alpha^5 }{2 n^3} \left( 1+ \frac{\alpha^2}{3 n^2}\right) \, ,
     \label{gamma_ann_S_wave_vec}
      \nonumber
    \\
\end{eqnarray}
and 
\begin{eqnarray}
  \Gamma_{\textrm{ann}}^{nP_J}((X\bar{X})_n \to \gamma \gamma)  = \frac{ |R'_{nP}(0)|^2}{\pi M^4}  {\rm{Im}}[f(^3 P_J)] = \begin{cases}
  \frac{M \alpha^7}{24 n^5} (n^2-1) \; , J=0 
  \\
   \frac{M \alpha^7}{90 n^5} (n^2-1) \; , J=2 
  \end{cases} \, ,
    \label{gamma_ann_P_wave_vec}
 \end{eqnarray}
where the bound-state wave functions and energy levels are taken at leading order, namely $|R_{nS}(0)|^2=4/(n^3 a_0^3)$ and $|R'_{nP}(0)|^2=4(n^2-1)/(9 n^5 a_0^5)$, with $a_0=2/M\alpha$ the Bohr radius. 
\begin{figure}[t!]
    \includegraphics[width=1.0\textwidth]{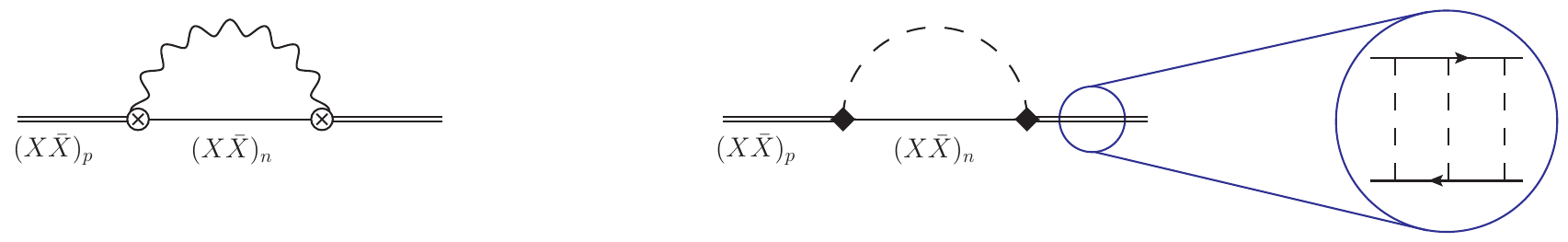}
    \caption{One-loop self-energy diagrams of a scattering state $(X\bar{X})_p$ in pNRQED$_{\textrm{DM}}$ (left) and pNRY$_{\gamma^5}$ (right). The internal lines are bound states, a vector mediator (wiggly line) and a scalar mediator (dashed line). The electric-dipole is shown with a cross vertex, whereas the quadruple with a diamond vertex. The zooming of the scattering state shows the actual content of the bilocal fields in pNREFTs, namely interacting pairs.}
    \label{fig:FrPr_fig1.pdf}
\end{figure}

The ultrasoft vertex governs the transitions among dark matter pairs. For a vector mediator and in pNRQED$_{\textrm{DM}}$, the leading term is  
the \emph{electric dipole interaction} of the dark fermion-antifermion pair with ultrasoft dark photons, see eq.~\eqref{pNREFT_vector_tot}, that comprise thermal photons as well. The dipole interaction is needed to compute the bound-state formation process $(X\bar{X})_p \to \gamma + (X\bar{X})_n$ and one can extract the corresponding cross section by taking the imaginary part of the self-energy diagram displayed in figure~\ref{fig:FrPr_fig1.pdf} (left diagram). In the thermal field theory version of pNRQED$_{\textrm{DM}}$, the inclusive bound-state formation cross section is (see ref.~\cite{vonHarling:2014kha} for the original derivation with a Bethe-Salpeter approach, and refs.~\cite{Binder:2020efn,Biondini:2023zcz} for recent derivations within pNREFTs)
\begin{align}
  (\sigma_{\hbox{\scriptsize bsf}} \, v_{\hbox{\scriptsize rel}})(\bm{p})\big\vert_{\textrm{vector}} =  \sum \limits_n   (\sigma^n_{\hbox{\scriptsize bsf}} \, v_{\hbox{\scriptsize rel}})(\bm{p}) 
  =\frac{4 \alpha}{3} \sum_{n} \left[ 1 + n_{\text{B}}(\Delta E_{n}^{p}) \right] |\langle n | \bm{r} | \bm{p}\,\rangle|^2  (\Delta E_{n}^{p})^3 \, .
\label{bsfn}  
\end{align}
The appearance of the Bose enhancement for the emitted mediator through the Bose-Einstein distribution comes naturally from pNRQED$_{\textrm{DM}}$ at finite temperature. 
The subscript will serve to distinguish the cross section from the corresponding one in the case of a scalar mediator (cfr.~eq.~\eqref{bsf_matrix_element_sca}). 
The energy splitting between a scattering state and a bound state, or equivalently the energy carried away from the emitted massless vector, is
\begin{equation}
\Delta E_{n}^{p} \equiv E_p-E_n = \frac{M}{4}v_{\textrm{rel}}^2\left(1+\frac{\alpha^2}{n^2v_{\textrm{rel}}^2}\right) \,,
\label{energy_photon_bsf}
\end{equation}
which holds at leading order.
As a reference, and for later comparison with the scalar-mediator case, we give the explicit expression of the bound-state formation of the ground state, which reads 
\begin{eqnarray}
(\sigma^{1\textrm{S}}_{\hbox{\scriptsize bsf}} v_{\hbox{\scriptsize rel}})(\bm{p}) \big\vert_{\textrm{vector}}=
    \frac{2^9}{3}\frac{ \pi \alpha^2}{M^2} S_0(\zeta) \frac{\zeta^4}{(1+\zeta^2)^2}  e^{-4 \zeta \arccot(\zeta)} 
      \, \left[ 1 + n_{\text{B}}(\Delta E_{1}^{p}) \right] \, .
     \label{our_gamma_1s} 
\end{eqnarray}
\subsection{Dark matter with a scalar mediator}
\label{sec:scalar_med_model}
 As in the previous model, we  assume the DM particle to be a Dirac fermion that carries no charge under the SM gauge group.  Dark matter fermions experience, however, an interaction that is mediated by a scalar particle of the hidden sector via Yukawa-type interactions.  The Lagrangian density of the model reads \cite{Pospelov:2007mp,Kaplinghat:2013yxa,Wise:2014jva}
 \begin{equation}
    \mathcal{L}= \bar{X} (i \slashed{\partial} -M) X + \frac{1}{2} \partial_\mu \phi \, \partial^\mu \phi -\frac{1}{2}m_\phi^2 \phi^2 -  \bar{X} (g + ig_5 \gamma_5)  X \phi - \frac{\lambda_\phi}{4!} \phi^4  +\mathcal{L}_{\hbox{\scriptsize portal}} \, ,
    \label{lag_mod_relativistic_sca}
    \end{equation}
 where $X$ is the DM Dirac field and $\phi$ is a real scalar. The scalar self-coupling is denoted with $\lambda_\phi$, whereas the scalar and pseudo-scalar couplings with the fermion are $g$ and $g_5$ respectively. For simplicity we assume the scalar self-coupling to be negligible and then plays no role in our analyses.\footnote{Such an interaction would be responsible for the generation of a thermal mass for the scalar mediator in the early universe, $m_{\textrm{thermal}} = T \sqrt{\lambda_\phi/12}$. Moreover, it induces bound-state formation via the emission of two scalar mediators \cite{Oncala:2018bvl}.}  The mass of the scalar mediator $m_\phi$ is assumed to be much smaller than the DM mass and, in order to compare the relevant observables with the gauge-invariant model where $m_\gamma=0$, we restrict to the situation $m_\phi \ll M \alpha^2$. To a good approximation, we can then treat the corresponding bound states as Coulombic. 
 
 In this work, we consider the case where the scalar coupling is larger than the pseudo-scalar coupling, namely $\alpha \equiv g^2/(4 \pi) \gg \alpha_5 \equiv g_5^2/(4 \pi)$. In so doing, we ensure that the dominant non-perturbative effects are originated from a scalar-type interaction, that induce an attractive potential, and we can largely neglect the mixed scalar-pseudoscalar and pure-pseudoscalar induced contributions \cite{Kahlhoefer:2017umn,Biondini:2021ycj}. The presence of pseudo-scalar interactions induces $S$-wave pair annihilation for this model, cfr.~eq.~\eqref{ann_cross_Sommerfeld_sca}.
Portal interactions are important for the model phenomenology and consistency. The scalar mediator of the dark sector couples to the SM via a Higgs-portal interaction, see e.g.~\cite{Arcadi:2019lka}. As in the vector mediator case, and for the sake of extracting the relic density, the details of the portal Lagrangian are not needed and we neglect the corresponding term in the following.

The low-energy theory that is obtained via a two-step matching from the model in eq.~\eqref{lag_mod_relativistic_sca}, and with the hierarchy of scales \eqref{H_scales}, results in a pNREFT-like Lagrangian \cite{Biondini:2021ccr,Biondini:2021ycj}
\begin{align}
    \mathcal{L}_{\hbox{\tiny pNRY$_{\gamma_5}$}}  &= \int d^3 \bm{r}   \, \varphi^\dagger(\bm{r},\bm{R},t) \left\lbrace  i \partial_0 +\frac{\bm{\nabla}^2_{\bm{r}}}{M} +\frac{\bm{\nabla}^2_{\bm{R}}}{4M} + \frac{\bm{\nabla}^4_{\bm{r}}}{4 M^3} - V(\bm{p},\bm{r},\bm{\sigma}_1,\bm{\sigma}_2)\right. \nonumber 
    \\
    & \left. - 2 g \phi(\bm{R},t) -g\frac{ r^i r^j }{4}  \left[ \nabla_R^i \nabla_R^j \, \phi (\bm{R},t)   \right] -   g \phi(\bm{R},t) \frac{\bm{\nabla}^2_{\bm{r}}}{M^2}    \right\rbrace \varphi(\bm{r},\bm{R},t) \nonumber
    \\
    &+  \frac{1}{2} (\partial^\mu \phi(\bm{R},t))^2  -  \frac{m_\phi^2}{2} \phi(\bm{R},t)^2 - \frac{ \lambda_\phi}{4!} \phi(\bm{R},t)^4  ,
    \label{pNREFT_sca_0}
 \end{align}
where the square brackets in the second line of eq.~\eqref{pNREFT_sca_0} indicate that the spatial derivatives act on the scalar field only, which is multipole expanded. It is worth mentioning the difference between the ultrasoft vertices of pNRY$_{\gamma^5}$ and pNRQED$_{\textrm{DM}}$. In the second line of eq.~\eqref{pNREFT_sca_0}, we see the appearance of a \textit{monopole} and a \textit{quadrupole} interaction as well as an interaction involving the derivative in the relative distance, whereas the dipole interaction is absent (see eq.~\eqref{pNREFT_vector_tot}). 

The annihilation of heavy DM pairs into scalar particles is described by the universal Lagrangian in eq.~\eqref{pNREFT_ann_generic}. One just has to obtain the specific matching coefficients for the model at hand. At leading order, the imaginary parts of the hard matching coefficients are \cite{Biondini:2021ycj}
\begin{eqnarray} 
       &&{\rm{Im}}[f(^1S_0)] = 2  \pi \alpha \alpha_5  \, , \quad  {\rm{Im}}[g(^1S_0)] = -\frac{8 \pi}{3} \alpha \alpha_5 \, , 
       \\
     &&  {\rm{Im}}[f(^3P_0)] =\frac{\pi}{6} (5 \alpha -  \alpha_5)^2 \, , \quad  {\rm{Im}}[f(^3P_2)] = \frac{\pi}{15}  (\alpha + \alpha_5)^2 \, .
\end{eqnarray}
They enable to extract the annihilation cross section
for scattering states and bound-state decay widths. The former observable reads  
 \begin{align}
     \sigma_{\textrm{ann}} v_{\textrm{rel}} = \frac{2 \pi \alpha \alpha_5}{M^2} \left( 1 - \frac{v^2_{\textrm{rel}}}{3}\right) S_0(\zeta)  + \frac{\pi (9 \alpha^2 - 2 \alpha \alpha_5 + \alpha_5^2)v_{\textrm{rel}}^2}{24 M^2} S_1(\zeta)    \, ,
     \label{ann_cross_Sommerfeld_sca}
 \end{align}
where the Sommerfeld factors are the same as the vector case, cfr.~eq.~\eqref{som_factors}, because they are obtained from the wave function of dark matter pairs that satisfy the same  Schroedinger equation with a Colomb potential. We notice that the psedoscalar interaction introduces a velocity independent $S$-wave contribution for the annihilation cross section in eq.~\eqref{ann_cross_Sommerfeld_sca}, which would be pure $P$-wave in the limit $\alpha_5 \to 0$. Projecting the operators of the Lagrangian in eq.~\eqref{pNREFT_ann_generic} onto bound states one obtains, as counterparts of the decay widths in eqs.~\eqref{gamma_ann_S_wave_vec} and \eqref{gamma_ann_P_wave_vec}, the following decays into scalar pairs 
 \begin{eqnarray}
  \Gamma_{\textrm{ann}}^{nS} = \frac{M  \alpha^4 \alpha_5}{n^3} \left( 1+ \frac{\alpha^2}{3 n^2}\right) \, , \quad \Gamma_{\textrm{ann}}^{nP_J}  = \begin{cases}
  \frac{M \alpha^5 (5 \alpha-\alpha_5)^2}{432 n^5} (n^2-1) \; , J=0 \\
   \frac{M \alpha^5 (5 \alpha-\alpha_5^2)}{1080 n^5} (n^2-1) \; , J=2 
  \end{cases} \, .
   \label{gamma_ann_S_and_P_wave_sca}
 \end{eqnarray}

The bound-state formation $(X\bar{X})_p \to \phi + (X\bar{X})_n$ is driven by the ultrasoft vertices of pNRY$_{\gamma^5}$. At variance with the vector mediator, the relevant interactions involve a quadrupole and a derivative vertex.  The monopole interaction, namely the first term in the second line of eq.~\eqref{pNREFT_sca_0}, does not contribute to the transitions between a scattering and a bound state because of the orthogonality of the corresponding wave functions. The bound-state formation cross section is extracted from the imaginary part of one-loop thermal self-energy diagrams, see exemplary diagram in figure~\ref{fig:FrPr_fig1.pdf} (right panel). The inclusive thermal cross section reads 
\begin{eqnarray}
(\sigma_{\textrm{bsf}} \, \vrel)(\bm{p}) \big\vert_{\textrm{scalar}} &=&  \alpha  \sum_{n} \left\lbrace \frac{ (\Delta E_n^p) ^5 }{120 }   \left[ |\langle\bm{p} | \bm{r}^2 |n  \rangle|^2 + 2 |\langle\bm{p} | r^i r^j |n  \rangle|^2\right] + 2   \, \Delta E_n^p    \Big \vert \Big \langle \bm{p} \Big \vert \frac{\nabla^2_{\bm{r}}}{M^2} \Big \vert n  \Big \rangle \Big \vert^{2} \right. \nonumber
 \\
  && \left.  -  \frac{(\Delta E_n^p)^3}{3}  \, \textrm{Re} \left[ \Big \langle \bm{p} \Big \vert \frac{\nabla^2_{\bm{r}}}{M^2} \Big \vert n  \Big \rangle  \langle n | \bm{r}^2 | p \rangle \right]  \, \, \right\rbrace  \left[ 1 + n_{\text{B}}(\Delta E_{n}^{p}) \right] \, ,
 \label{bsf_matrix_element_sca}
\end{eqnarray}
where three different matrix elements appear, at variance with the sole dipole matrix element in the vector case. Owing to the power counting offered by the pNREFT, and comparing the matrix elements, one can already notice that the bound-state formation cross section in eq.~\eqref{bsf_matrix_element_sca}  is $\alpha^2$ suppressed with respect to the case of a vector mediator. The bound state formation for the ground state is 
\begin{equation}
(\sigma^{1\textrm{S}}_{\hbox{\scriptsize bsf}} v_{\hbox{\scriptsize rel}})(\bm{p})  \big\vert_{\textrm{scalar}} = \frac{\pi \alpha^4}{M^2} S_0(\zeta)  \frac{2^6}{15}\frac{\zeta^2 (7 + 3 \zeta^2)}{(1+\zeta^2)^2} e^{-4 \zeta \arccot(\zeta)}   \, \left[ 1 + n_{\text{B}}(\Delta E_{1}^{p}) \right]  \, . 
    \label{bsf_1S}
    \end{equation}
\section{Thermal masses and dark matter freeze-in via decays}
\label{sec:thermal_masses_FI}
As anticipated in the introduction, the production of dark matter via freeze-in involves temperatures larger than and of the order of the dark matter mass \cite{McDonald:2001vt,Hall:2009bx} (see also \cite{Bernal:2017kxu} for a recent review). This calls for a careful scrutiny of thermal effects. A common modification for particles in a high-temperature environment is the appearance of thermal masses, which have only recently been addressed in the context of DM~\cite{Baker:2016xzo,Baker:2017zwx,Dvorkin:2019zdi,Darme:2019wpd,Chakrabarty:2022bcn,Bringmann:2021sth}. In these studies, the effect of thermal masses has been explored in decay processes which would be forbidden at zero temperature and instead open up in a thermal plasma, and in combination with phase transitions in the early universe. In this paper, we consider a situation where thermal masses can either suppress or enhance the decay process that sources the DM production. At variance with the effects that we have discussed in section~\ref{sec:NT_effects}, which increase particle interaction rates, here we focus on a situation where the particle production is going to be less efficient due to thermal masses.  We discuss an exemplary class of models where this situation occurs in the next section.
\subsection{Majorana dark matter and $t$-channel mediators}
\label{sec:model_t_channel}
Due to the increasing elusive character of the dark matter particle, it is well justified to assume it to be a SM gauge singlet. Out of the different portal realizations, a quite rich phenomenology is offered by DM coupled to the visible sector via an accompanying partner of the dark sector. The latter is taken to be charged under some (or all) gauge groups of the SM and, hence, it triggers experimental prospects for direct and indirect detection, as well as collider searches. This class of models is often referred to as $t$-channel mediator models \cite{Garny:2015wea,DeSimone:2016fbz,Arina:2020udz}. Here, the mediator stands for the particle of the dark sector that links the actual dark matter with SM fields, and it is not understood as a mediator of long-range interactions as in the former section~\ref{sec:NT_effects}. The dark matter particle and the mediator carry a $Z_2$ charge (they are odd under this symmetry), which makes the DM candidate stable in the first place upon assuming that it is the lightest state of the dark sector.
Dark matter can be either a scalar or a fermion and, depending on the SM fermion it interacts with, the gauge quantum numbers for the mediator can be fixed (see \cite{Garny:2015wea} and \cite{Belanger:2018sti}). 

In the following we shall consider two models where the DM is a Majroana fermion that interacts with (i) a right-handed up quark or (ii) a right-handed lepton. In both cases the mediator is a scalar, respectively either a triplet or a singlet under QCD, and we indicate it with $\eta$. The Lagrangian can be written as follows 
\begin{eqnarray}
 \mathcal{L} & = & 
 \mathcal{L}^{ }_{\hbox{\tiny SM}} + 
 \frac{1}{2} \, \bar{X} \left(  i \slashed{\partial} - M \right)  X 
 + (D^{ }_\mu \eta)^\dagger D^\mu \eta 
 - M_\eta^2\, \eta^\dagger \eta 
 - \lambda^{ }_2 (\eta^\dagger \eta)^2 
 \nonumber 
 \\
 & - & \lambda^{ }_3\, \eta^\dagger \eta\, \mathcal{H}^\dagger \mathcal{H} 
 - y\,  \eta \bar{X} P_R f 
 - y^* \eta^\dagger \, \bar{f} P_L X 
 \;,
 \label{Lag_gen_FI}
\end{eqnarray}
where $P_{L(R)}$ are the chiral projectors, $X$ is the Majorana fermion dark matter, $f=q,\ell$ is a SM fermion, $\mathcal{H}$ is the SM Higgs doublet, $M_\eta$ the mass of the mediator and $M$ the mass of the DM particle, with $M_\eta> M$ so to ensure a stable dark matter candidate. In the context of minimal flavour violation, we consider the coupling with one SM fermion generation at a time.  The hypercharge of the $\eta$ particle is then fixed to be $Y_\eta=-Y_f$. The covariant derivative comprises the corresponding relevant gauge fields (only $B_\mu$ for the interaction with a lepton and both $B_\mu$ and $A^{a}_\mu$, with $a=1,..,8$ for the interaction with right-handed quarks).\footnote{It is worth mentioning that thermal masses have been considered in the production of heavy neutrinos from charged scalar decays in ref.~\cite{Drewes:2015eoa} and in the context of see-saw type I leptogenesis in refs.~\cite{Anisimov:2010gy,Ghiglieri:2016xye}. Despite the models are different with respect to phenomenology, the topology of the relevant diagrams is indeed quite similar.}

In the freeze-in scenario the DM fermion only appears in the final state of the relevant  processes. DM production occurs via $1 \to 2$ decays in this model, namely $\eta \to X f$ and its complex conjugate, as well as through  $2 \to 2$ scatterings with SM particles. The latter are especially relevant for a compressed mass spectrum $\Delta M \equiv M_\eta-M  \ll M$. We remark here that  additional thermal phenomena can occur, and that they have been more carefully investigated for leptogenesis \cite{Anisimov:2010gy,Ghisoiu:2014mha,Ghiglieri:2016xye}, such as multiple soft scatterings, namely the LPM effect. The latter typically increases the production rate of a feeble interacting particle via effective $1 +n \leftrightarrow 2 + n$ processes, see \cite{Biondini:2020ric} for their inclusion in the context of freeze-in DM. In this work, we aim to highlight the subset of thermal effects as restricted to thermal masses, which can make the production rate smaller. Such an approach already goes beyond the standard treatment in the present literature, where $1 \to 2$ decays are estimated with \textit{in-vacuum} masses \cite{Garny:2018icg,Garny:2018ali,Belanger:2018sti,Junius:2019dci,Bollig:2021psb}.
\begin{figure}
    \centering
    \includegraphics[width=1.0\textwidth]{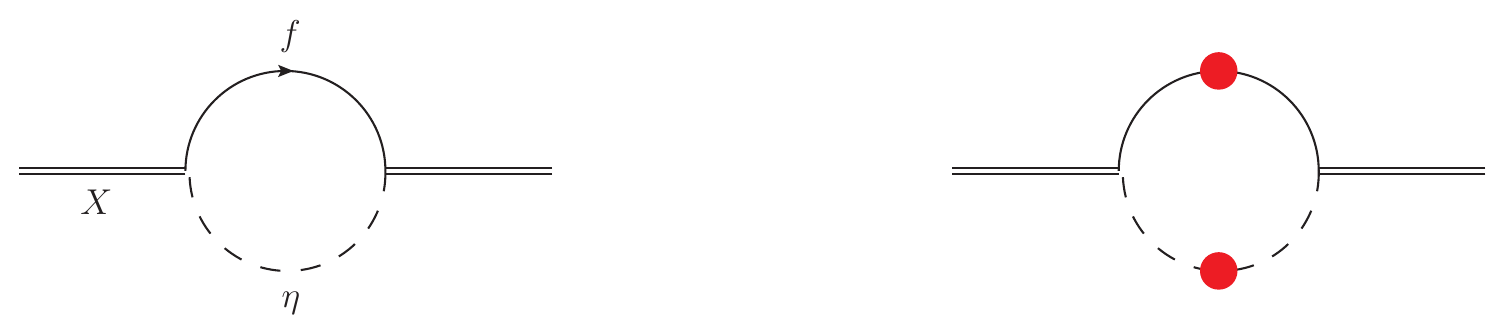}
    \caption{One-loop self-energy diagrams for the Majorana fermion dark matter $X$ (solid double line). The scalar mediator and the SM fermion are displayed with dashed and solid lines respectively. The left diagram stands for a scalar mediator and SM with in-vacuum masses, whereas resummed propagators with thermal masses are shown in the right diagram.}
    \label{fig:FrPr_fig2.pdf}
\end{figure}

We shall derive the DM production rate in the formalism offered by the spectral function. For practical calculations, the spectral function of the produced particle can be related to the imaginary part of its retarded correlator at finite temperature $\textrm{Im} \Pi_R$ \cite{Asaka:2006rw,Bodeker:2015exa,Laine:2016hma}. At leading order, one has to compute the imaginary part of the thermal one-loop self-energy of the dark matter particle, see figure~\ref{fig:FrPr_fig2.pdf}. The main advantage of such a formalism is that one can easily generalize a given process to higher order and include thermal effects.  
The imaginary part of the retarded correlator enters the rate equation that governs the evolution of the DM particle. As we are in the freeze-in scenario, there is no loss term, and the rate equation is \cite{Bodeker:2015exa,Biondini:2020ric}
\begin{eqnarray}
\dot{n}_{X}+ 3H n_{X} &=& 2 |y|^2 \int_{\bm{k}} \frac{n_\mathrm{F}(k^0)}{k^0}  \textrm{Im} \Pi_R \, ,
\label{number_density_Born}
\end{eqnarray}
where $\int_{\bm{k}}\equiv \int d^3k/(2\pi)^3$ and the factor of two counts the helicity states of the Majorana fermion. 
We consider only $1 \to 2$ decays as contributing to $\textrm{Im} \Pi_R$, however, we assess the modification that may occur when including thermal masses.

In the high-temperature limit, that corresponds to temperatures larger than any in-vacuum mass scale, repeated interactions with the plasma constituents generate the so-called \textit{asymptotic masses}. For the scalar mediator, SM right-handed quarks and leptons, they read \cite{Kajantie:1995dw,Giudice:2003jh,Biondini:2018pwp}
\begin{eqnarray}
&&\bar{X} \eta P_R q: \;\;  m^2_\eta= \left( \frac{g_3^2C_F+Y_q^2g_1^2}{4} + \frac{\lambda_3}{6}  \right)  T^2 \, , \quad m_q^2= \frac{ T^2}{4}(g_3^2C_F + Y_q^2g_1^2 + |h_q|^2) \, ,
\label{asym_the_mass_q}
\\
&&\bar{X} \eta P_R \ell: \;\;  m^2_\eta= \left( \frac{Y_\ell^2g_1^2}{4} + \frac{\lambda_3}{6}  \right)  T^2 \, , \quad m_q^2= \frac{ T^2}{4}(Y_\ell ^2g_1^2 + |h_\ell|^2) \, ,
\label{asym_the_mass}
\end{eqnarray}
where  $C_F=(N_c^2-1)/(2N_c)$ is the quadratic Casimir of the fundamental 
representation, $g_1$ and $g_3$ are the SM U(1)$_Y$ and SU(3) gauge couplings, and $h_f$ is the Higgs-fermion Yukawa coupling. We distinguish the two models explicitly and show the corresponding thermal masses $m_i$ (capital letters are instead used to indicate the in-vacuum masses). The thermal mass for the DM is negligible since it is proportional to $|y|^2 \ll g_3^2, g_1^2, \lambda_3,|h_q|^2$; for the freeze-in production to be applicable the coupling of the DM with other particles is $y \lesssim \mathcal{O}(10^{-8})$ \cite{McDonald:2001vt,Hall:2009bx}. As we shall restrict to temperatures above the electroweak scale, the thermal masses for the right-handed fermions is the only source of a mass term.\footnote{For temperatures smaller than about $T_c \simeq 150$ GeV, the SM Higgs boson undergoes a crossover, then the top-quark and the mediator $\eta$ thermal masses would acquire a more complicated form. Despite this aspect is interesting, it is not expected to qualitatively change the effect on the DM production, and we leave it for future investigations. } This is not true for the scalar particle $\eta$: the asymptotic thermal mass of the scalar is not a good approximation when the vacuum mass $M_\eta$ is no longer negligible  with respect to thermal scales. As soon as 
$M_\eta\gtrsim  T$, one must  include it in the determination of the thermal self-energy of the scalar, which in turn gives the thermal contribution to the thermal mass $m_\eta$. With minimal modifications, we can adapt the result from ref.~\cite{Biondini:2020ric}, and include an accurate temperature dependence for the scalar mediator mass, which decomposes in an in-vacuum and thermal contributions $\mathcal{M}_\eta^2 \equiv M_\eta^2 + m_\eta^2$. The thermal mass $m_\eta^2$ for $T \sim M_\eta$ can be found in ref.~\cite{Biondini:2020ric} for the interaction with SM quarks.

We can now proceed with the calculation of the thermal process $\eta \to X f$, which is sometimes referred to as Born rate because it corresponds to the leading process that drives the freeze-in production. We present the result with finite thermal masses first, which corresponds to the right diagram in figure~\ref{fig:FrPr_fig2.pdf} (red bubbles stand for resummed propagators with thermal masses). Then, we show the in-vacuum limit $m_i=0$, which originates from the left diagram in figure~\ref{fig:FrPr_fig2.pdf}. The production rate reads 
\begin{equation}
\label{fullborneta}
    {\rm{Im}}\Pi_{\hbox{\tiny R},\eta \to X f}=\frac{N^{f}_{c}}{16\pi k}\int_{p_\mathrm{min}}^{p_\mathrm{max}} d p [\mathcal{M}_\eta^2-M^2-m_f^2-2 k^0 (E_p-p)]  [n_\mathrm{B}(k^0+E_p)+n_\mathrm{F}(E_p)],
\end{equation}
where we have explicitly indicated that we single out the process $\eta \to X f$. Then, $N_{c}^q=3$ for a quark and $N_{c}^{\ell}=1$ for a lepton, $E_p=\sqrt{p^2+m_f^2}$ and the integration boundaries are
\begin{equation}
    p_\mathrm{min,\,max}=\frac{\mathcal{M}_\eta^2-M^2-m_f^2}{2M^2}\left|k^0\sqrt{1-\frac{4M^2 m_f^2}{(\mathcal{M}_\eta^2-M^2-m_f^2)^2}}\mp k\right|.
    \label{boundaries_thermal_born}
\end{equation}
It is useful to perform the in-vacuum mass limit, which gives an analytical expression for the Born rate, and will serve as a reference for the corresponding result with thermal masses. It reads 
\begin{eqnarray}
    && {\rm{Im}}\Pi_{\hbox{\tiny R},\eta \to X f} \big\vert_{m_i=0} = \frac{N^f_c (M_\eta^2 -M^2)}{16\pi k} \int_{p_{\hbox{\tiny min}}}^{p_{\hbox{\tiny max}}} dp \left[ n_\mathrm{B}(p+k^0) + n_\mathrm{F}(p)\right]  \nonumber
     \\
     &&~~~~~~=
 \frac{N^f_c T (M_\eta^2 -M^2)}{16 \pi k} \left[ \ln \left( \frac{\sinh(\beta(k^0+p_{\hbox{\tiny max}})/2)}{\sinh(\beta(k^0+p_{\hbox{\tiny min}})/2)}\right) - \ln \left( \frac{\cosh(\beta p_{\hbox{\tiny max}} /2)}{\cosh(\beta p_{\hbox{\tiny min}}/2)}\right)\right] \, , 
 \label{vacuum_born}
 \end{eqnarray}
 where 
 \begin{equation}
     p_{\hbox{\tiny min}}=\frac{M^2_\eta - M^2 }{2(k^0+k)} \, , \quad p_{\hbox{\tiny max}}=\frac{M^2_\eta - M^2 }{2(k^0-k)} \, .
     \label{boundaries_vacuum_born}
 \end{equation}
 \begin{figure}[t!]
    \centering
    \includegraphics[width=0.47\textwidth]{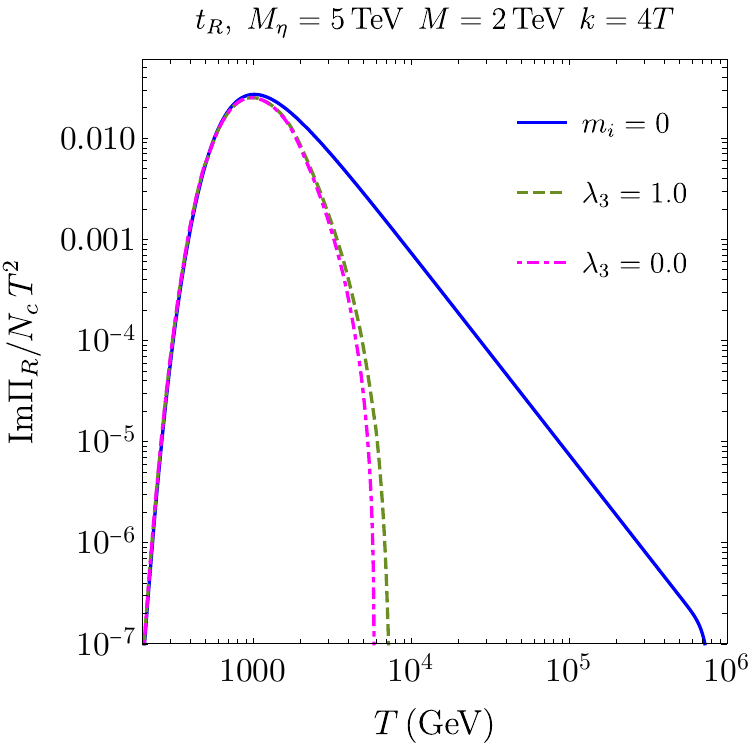}
    \hspace{0.2 cm}
      \includegraphics[width=0.47\textwidth]{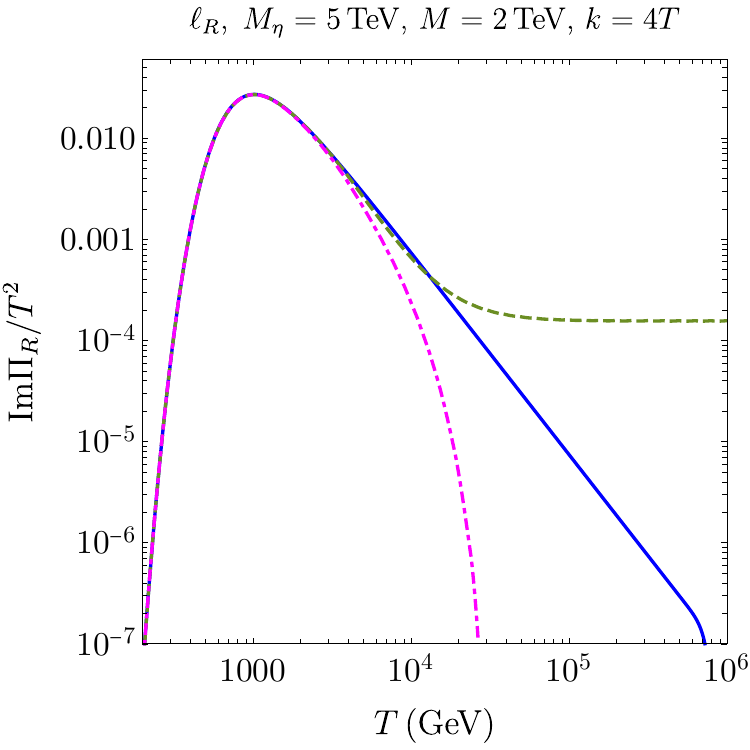}
    \caption{The production rate $\textrm{Im}\Pi_R/N_c^fT^2$ for $\eta \to X f$ decay processes as function of the temperature. The left panel is for a right-handed top quark, the right panel for a right-handed lepton. The production rates are taken for a fixed value of the DM momentum $k=4T$.}
    \label{fig:IMR_rate_l_and_q}
\end{figure}
 The Born rate is our key ingredient to extract the DM energy density, cfr.~eq.~\eqref{number_density_Born}, and to assess the interplay with modified cosmological histories. The corresponding numerical results are collected in  section~\ref{sec:num_FI}. In figure~\ref{fig:IMR_rate_l_and_q} we show the production rate $ {\rm{Im}}\Pi_{\textrm{R}}$ for the top-quark case (left panel) and a right-handed lepton (right panel). By choosing the top quark we can inspect the impact of a finite Yukawa coupling $h_f$ in eq.~\eqref{asym_the_mass_q}, which would be negligible for the other quarks and all the leptons. The in-vacuum masses of the dark sector particles, $M=2$ TeV and $M_\eta=5$ TeV, are chosen such that the freeze-in contribution to $\Omega_{\textrm{DM}}h^2$ is largely dominant with respect to the one from the super-WIMP mechanism (see \cite{Garny:2018ali,Biondini:2020ric}).\footnote{In this class of modes the super-WIMP mechanism comes along the freeze-in. The former takes place at much smaller temperatures $T \ll M_\eta$ and the relevant process is the freeze-out of the mediator $\eta$ and its subsequent late decays into DM particles.} For this choice of the masses, the $2 \to 2$ scatterings are also moderate with respect to the DM production from decays (see refs.~\cite{Garny:2018ali,Biondini:2020ric}). For a top-philic DM, the inclusion of thermal masses always suppress the production rate with respect to the in-vacuum result (solid-blue line). Moreover, different values of the portal coupling $\lambda_3 \in [0,1]$ have a rather marginal impact on $ {\rm{Im}}\Pi_{\textrm{R}}$. A different effect due to thermal masses is instead found in the lepton case. Here, the outcome holds equally for each family since $h_\ell \ll 1$ for electrons, muons and taus. The thermal correction to $m_\eta$ that is proportional to $\lambda_3$, see eq.~\eqref{asym_the_mass}, plays a more important role and it can make the production rate even larger than the in-vacuum limit (at high temperatures the two-body phase space is increased by a large $m_\eta$). For $\lambda_3 = 0$ one finds again a suppressed rate with respect to the vacuum masses, though the Born rate with in-vacuum and thermal masses are closer with respect to the top-philic scenario.   In section~\ref{sec:num_FI} we focus on scenarios where thermal masses inhibit DM production with respect to the in-vacuum mass limit. We then fix $\lambda_3=0$, which reduces the number of free parameters of the models, while preserving their rich phenomenology.\footnote{An important and relevant consequence for $\lambda_3 \sim \mathcal{O}(1)$ is the possibility to trigger a first order electroweak phase transition for the lepton scenario, see refs.~\cite{Liu:2021mhn,Biondini:2022ggt}.}
\section{Modified Cosmological histories}
\label{sec:mod_cosmo}
The extraction of the DM energy density depends on the thermal history of the universe via the Hubble rate. This is the clock that measures the efficiency of a given particle rate in an expanding background. The standard procedure is to assume that the DM freeze-out and freeze-in occur in an epoch of radiation domination, where the SM is the dominant component at temperatures $T \gg \mathcal{O} (1) \, \textrm{MeV}$. However, there are no obvious reasons for limiting ourselves on such cosmological history. The paradigm of inflationary cosmology calls for a stage of reheating that goes along with well motivated different expansion histories, which include early matter dominated phase, moduli fields and quintessence fluids \cite{Allahverdi:2010xz,Berlin:2016vnh,Tenkanen:2016jic,Berlin:2016gtr,Dine:1995uk,Caldwell:1997ii,Sahni:1999gb}. The latter option has ties with the current accelerated expansion of the universe. 
In this paper, we do not select a particular ultraviolet completion and we consider a family of modified cosmological histories that provides a faster expansion rate before BBN. More specifically, we follow the framework proposed in ref.~\cite{DEramo:2017gpl}. 
\subsection{A faster universe expansion}
Following the approach in ref.~\cite{DEramo:2017gpl}, a modification of the universe expansion history is achieved by introducing another species $\varphi$, that redshifts as $\varphi \propto a^{-(4+n)}$, where $a$ is the universe scale factor.\footnote{The same symbol is also used to indicate the bilocal field of the low-energy pNREFTs in section~\ref{sec:NT_effects}. Here $n$ enters the exponent of the scale factor, whereas it was used to label bound states in section~\ref{sec:NT_effects}.} For $n>0$ the energy density of $\varphi$ dominates the radiation component at early times, while it becomes completely negligible at later times. We label the corresponding energy densities as $\rho_\varphi$ and $\rho_{\textrm{rad}}$ respectively. In order to be quantitative on the relative importance of the additional fluid during the thermal history, one has to choose some reference temperature. We use the prescription  given in ref.~\cite{DEramo:2017gpl}, and we take the \emph{reference temperature} $\Tref$ as the temperature at which $\rho_{\varphi}=\rho_{\textrm{rad}}$. On general grounds, the smaller $\Tref$ the longer the faster expansion takes place and modifies the standard picture. A variety of cosmological scenarios are accounted for by two parameters $(\Tref,n)$. For example, the case $n=2$ describes the quintessence scenario ~\cite{Caldwell:1997ii,Sahni:1999gb}, which is also motivated by the discovery of the accelerated universe expansion. Such a case is also referred to as \emph{kination} regime, where the kinetic energy of the fluid is indeed dominant. Alternative realizations to the quintessence option, that still have the same redshift behaviour, are described in refs.~\cite{Wetterich:1987fm,Dimopoulos:2017zvq}. Examples of cosmological scenarios with $n>2$ are found in refs.~\cite{DEramo:2017gpl,Kamenshchik:2001cp,Chavanis:2014lra,Khoury:2001wf}. For our purpose,  the main point to be made is that, for temperature larger than $\Tref$, the universe expands \emph{faster}. Consequently the predicted dark matter energy density may change because the particle interaction rates becomes less effective.\footnote{In a recent work \cite{Davoudiasl:2023khm}, it has been considered the option where the universe expands slower than the standard cosmological history, and its impact on the DM relic density is discussed.}  

In order to provide a self-contained discussion, we streamline here the derivation of the main quantities that we need for the numerical extraction of the DM energy density. 
The steps for defining a modified effective number of relativistic degrees of freedom are as follows. First, one has consider the conservation of the total entropy and to assume that the standard radiation and additional specie $\varphi$ dominate the universe energy budget, namely $ \rho= \rho_{\textrm{rad}}+ \rho_{\varphi}$. This way one can express the ratio of the energy density $\rho_\varphi \propto a^{-(4+n)}$ at two different temperatures, which reads
\begin{equation}
    \frac{\rho_\varphi(T)}{\rho_\varphi(\Tref)} = \left( \frac{h_{\textrm{eff}}(T)}{h_{\textrm{eff}}(\Tref)} \right)^{\frac{4+n}{3}} \left( \frac{T}{\Tref} \right)^{4+n} \, ,
    \label{rho_phi_1}
\end{equation}
and use this condition to express the energy density $\rho_\varphi$ with the radiation temperature.  The total energy density can be written as follows 
\begin{equation}
   \rho(T)= \frac{\pi^2 T^4}{30} \left[ g_{\textrm{eff}}(T) + g_{\textrm{eff}}(\Tref) \left( \frac{h_{\textrm{eff}}(T)}{h_{\textrm{eff}}(\Tref)} \right)^{\frac{4+n}{3}} \left( \frac{T}{\Tref} \right)^{n} \right]  \, ,
   \label{rho_phi_2}
\end{equation}
where we have used the definition $\rho_\varphi(\Tref)= \rho_{\textrm{rad}}(\Tref)$ and $g_{\textrm{eff}}(T)$ are the temperature-dependent number of relativistic degrees of freedom. Reading off eq.~\eqref{rho_phi_2}, one can define a generalized number of relativistic degrees of freedom for the modified cosmologies $(\Tref,n)$
\begin{eqnarray}
    g^\varphi_{\textrm{eff}}(T,\Tref,n) \equiv  g_{\textrm{eff}}(T) + g_{\textrm{eff}}(\Tref) \left( \frac{h_{\textrm{eff}}(T)}{h_{\textrm{eff}}(\Tref)} \right)^{\frac{4+n}{3}} \left( \frac{T}{\Tref} \right)^{n}  \, .
       \label{rho_phi_3}
\end{eqnarray}
Let us remark that the limit to the standard cosmology is not recovered by setting $n=0$. Rather, in this case, one has a double copy of a radiation-like fluid, which gives a factor of 2 for $T=\Tref$ in eq.~\eqref{rho_phi_3}. 
The Hubble rate that enters the relevant Boltzmann equations in section~\ref{sec:numerics} then reads 
\begin{equation}
    H^{\varphi}= \sqrt{\frac{4 \pi^3}{45} \,   g^\varphi_{\textrm{eff}}(T,\Tref,n)  } \, \frac{T^2}{M_{\textrm{Pl}}} \, ,
    \label{mod_hubble_rate}
\end{equation}
where $M_{\textrm{Pl}} \simeq 1.22 \times 10^{19}$ GeV is the Planck mass.
Before concluding the section, let us briefly recall that an unavoidable constraint on the energy density of the additional field/fluid $\varphi$ has to be imposed. Indeed, a faster expansion rate has to be limited at times (or temperatures) before the Big Bang Nucleosynthesis, which occurs at $T_{\textrm{BBN}} \simeq 4$ MeV \cite{Kamenshchik:2001cp,Hannestad:2004px}. The remarkable agreement between the predictions and  measurements of light element abundances sets a cornerstone of particle cosmology. If the universe expands too fast at around the BBN epoch, then the light elements could not even form or, in any case, their abundance would sensibly change.  The effect of the additional component $\varphi$ is parameterized in terms of an effective number of relativistic degrees of freedom, more specifically adding up to the number of effective neutrinos, and we adopt here a limit on the reference temperature  namely $\Tref \geq (15.4)^{1/n}$ MeV \cite{DEramo:2017gpl}. We note in passing that, by considering DM candidates with masses $M \gtrsim \mathcal{O}(100)$ GeV, reference temperatures quite larger than such lower bound are sufficient to highlight the effect of modified cosmologies. In section~\ref{sec:numerics} we will not consider reference temperatures smaller than the QCD crossover, namely $\Tref \geq 154$ MeV~\cite{HotQCD:2014kol}. 
\section{DM energy density}
\label{sec:numerics}
In this section, we combine the improved interaction rates that were obtained in sections~\ref{sec:NT_effects} and \ref{sec:thermal_masses_FI} with the  modified cosmological histories of section~\ref{sec:mod_cosmo}. Our aim is to  quantitatively show the interplay of a faster expansion rate of the universe with (i) larger cross sections from non-perturbative effects for the freeze-out scenario;  (ii) a reduced particle production due to thermal masses in the case of freeze-in. 
\subsection{Freeze-out}
\label{sec:num_FO}
In order to capture the DM annihilation in the form of bound states, we rely upon an  effective description, which is commonly adopted in the literature and was originally proposed in ref.~\cite{Ellis:2015vaa}. In the most general case, the situation is rather complex because there is an equation for the DM particle number density, denoted by $n_{ X}$, and an equation for the number density of each bound state. A network of coupled  Boltzmann equations would then need to be solved. However, whenever the reactions that drive the rate of change of the bound states are faster than the Hubble rate, the network of Boltzmann equations for the bound states significantly simplifies and turns into algebraic equations \cite{Ellis:2015vaa}.\footnote{Such treatment has been very recently revisited in \cite{Garny:2021qsr,Binder:2021vfo} to include transitions among bound states, which  make bound-state effects more prominent.}  The relevant particle rates are the bound state dissociation rate and the bound state decay width, which are both much larger than the Hubble rate for the mass parameters and couplings that we consider in this work.  We have carefully checked that, even in the case of a faster universe expansion of section~\ref{sec:mod_cosmo}, these conditions hold for the considered masses and couplings.
As a result of such approximations, a single Boltzmann equation for $n_X$ is found, where the reprocessing of fermion-antifermion pairs into bound states and their decays is accounted for by an effective cross section
\begin{equation}
    \frac{d n_{X}}{d t} + 3H n_{X} = - \frac{1}{2}\langle \sigma_{\textrm{eff}} \, v_{\textrm{rel}} \rangle (n^2_{X}-n^2_{X,\textrm{eq}}) \, .
    \label{Boltzmann_eq_eff}
\end{equation}
The factor of $1/2$ on the right-hand side of eq.~\eqref{Boltzmann_eq_eff} appears because we consider Dirac DM particles  hence not self-conjugated) \cite{Gondolo:1990dk}.
For the Hubble rate we will adopt the one in a standard cosmological history, namely $H=\sqrt{8 \pi e /3}/M_{\textrm{Pl}}$, with the radiation energy density $e=\pi^2 T^4 g_{\textrm{eff}}/30$, as well as the more general expression in eq.~\eqref{mod_hubble_rate}, which accounts for a faster expansion before the BBN epoch. The SM contribution to $g_{\textrm{eff}}$ is taken from ref.~\cite{Laine:2015kra}.   The effective thermally averaged cross section, upon neglecting  bound-to-bound transitions, yields
\begin{equation}
    \langle  \sigma_{\textrm{eff}} \, v_{\textrm{rel}} \rangle  =  \langle \sigma_{\textrm{ann}} \, v_{\textrm{rel}} \rangle + \sum_{n} \langle   \sigma^n_{\textrm{BSF}} \, v_{\textrm{rel}} \rangle \, \frac{\Gamma_{\textrm{ann}}^n}{\Gamma_{\textrm{ann}}^n+\Gamma_{\textrm{BSD}}^n} \, , 
    \label{cross_section_eff_BE}
\end{equation}
where the sum runs over all bound states. The thermal averaging is carried out in the standard way, i.e.~we take Maxwell Boltzmann distributions for the incoming DM fermion and antifermion \cite{Gondolo:1990dk,vonHarling:2014kha}. The Sommerfeld corrected annihilation cross sections and the BSF cross section are given in eqs.~\eqref{ann_cross_Sommerfeld_vec} and \eqref{bsfn} for the vector mediator, whereas in eqs.~\eqref{ann_cross_Sommerfeld_sca} and \eqref{bsf_matrix_element_sca} for the scalar mediator. The decays widths $\Gamma^n_{\textrm{ann}}$ are found in eqs.\eqref{gamma_ann_S_wave_vec}, \eqref{gamma_ann_P_wave_vec} and \eqref{gamma_ann_S_and_P_wave_sca}. The bound-state dissociation width, which corresponds to the breaking of the bound-state via the absorption of a mediator from the thermal bath, can be obtained from the bound-state formation cross section via detailed balance~\cite{vonHarling:2014kha}, also known as Milne relation for the particular case. In the following, we will consider excited states up to $n\leq 2$, hence we include the bound states 1S, 2S and 2P (the latter state comprises three states for the magnetic quantum number degeneracy). 
\begin{figure}[t!]
    \centering
    \includegraphics[scale=0.55]{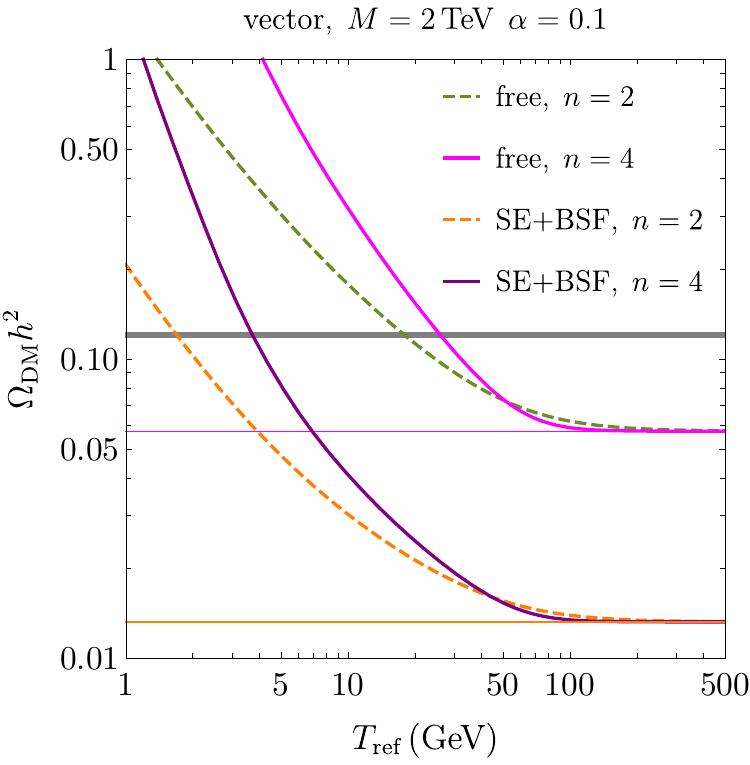}
    \hspace{0.5 cm}
    \includegraphics[scale=0.55]{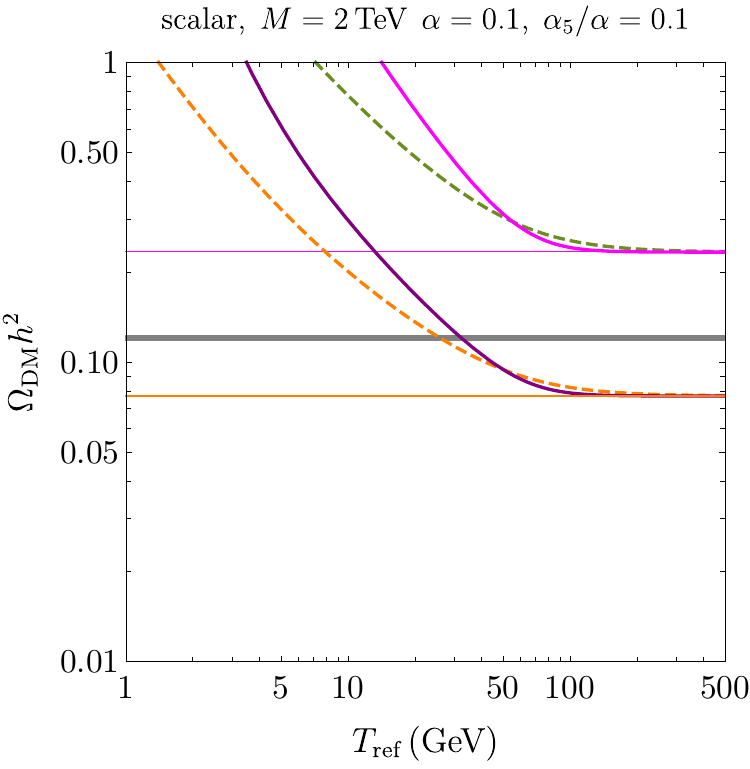}
    \caption{Dark matter energy density for the vector and scalar mediator models, respectively left and right panel, as function of the reference temperature $\Tref$.  The coupling among the DM fermion and the mediator is set to $\alpha=0.1$ and $M=2$ TeV (for the scalar-mediator model $\alpha_5/\alpha=0.1$). }
    \label{fig:Omega_versus_Tr}
\end{figure}

We shall present some numerical results where we aim to highlight the combination of larger particle rates and a modified cosmological history. The same version of each plot is presented for the two benchmark models, respectively DM fermion with a vector and scalar mediator as discussed in sections~\ref{sec:abelian_model} and \ref{sec:scalar_med_model}. We start with figure~\ref{fig:Omega_versus_Tr}, where we display the DM energy density as  function of the reference temperature $\Tref$. The DM mass and the couplings are indicated at the top label of each plot. The minimal reference temperature is well above the lower bounds $\Tref^{\textrm{min}} \simeq 3.9, 1.9$ MeV for $n=2$ and $n=4$ respectively. The dashed-green and dashed-orange lines stand for the DM energy density as obtained with free and improved cross sections for the kination option $n=2$, whereas solid magenta and purple lines correspond to an alternative cosmology with $n=4$. As a general common feature, one may notice how the DM energy density converges to the values that are obtained in the standard cosmology  (solid-thin horizontal lines) for large enough reference temperatures, here $\Tref \gtrsim 100$ GeV. This is traced back to the freeze-out happening at temperatures where the standard expansion rate is recovered. The situation changes substantially for smaller $\Tref$'s, which progressively make a faster expansion last longer.  The annihilation cross section becomes less effective and, therefore, larger DM abundance are found. This holds irrespective of free or improved cross sections, where non-perturbative effects are included.  For the specific choice of the mass and couplings, it is worth noticing that modified cosmologies open a window for $\Tref$ where the observed DM energy density $\Omega_{\textrm{DM}}h^2 = 0.012 \pm 0.012$ is reproduced. Indeed, both for the vector and scalar mediator models with a standard cosmological history,  the predicted energy density is either a fraction of the observed value or would overclose the universe (see solid-thin lines). The pseudoscalar coupling is fixed to $\alpha_5=0.1 \alpha$ in the following (see ref.~\cite{Biondini:2023ksj} for a more detailed study on the dependence of near-threshold effects with varying $\alpha_5$). 
\begin{figure}[t!]
    \centering
    \includegraphics[width=0.47\textwidth]{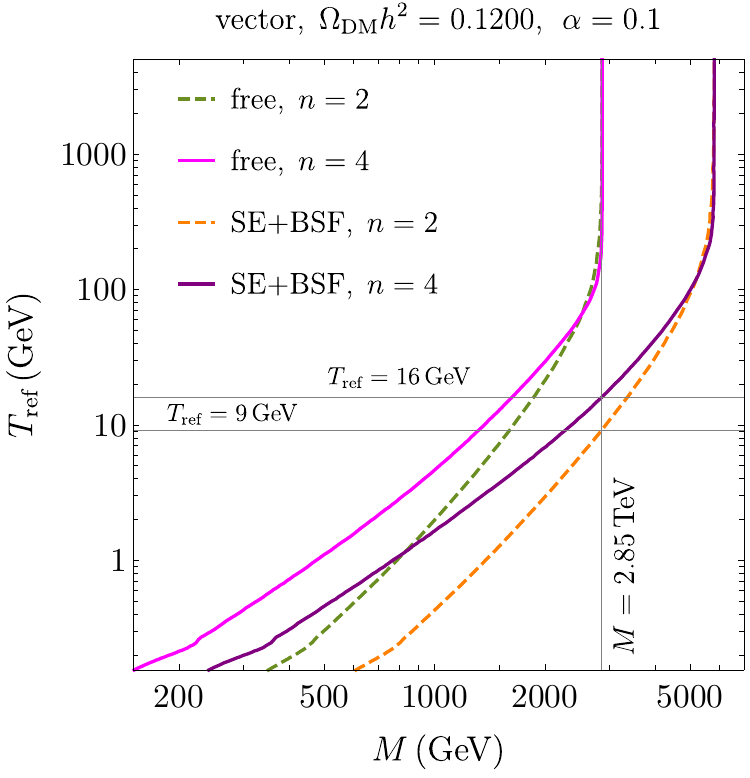}
    \hspace{0.2 cm}
    \includegraphics[width=0.47\textwidth]{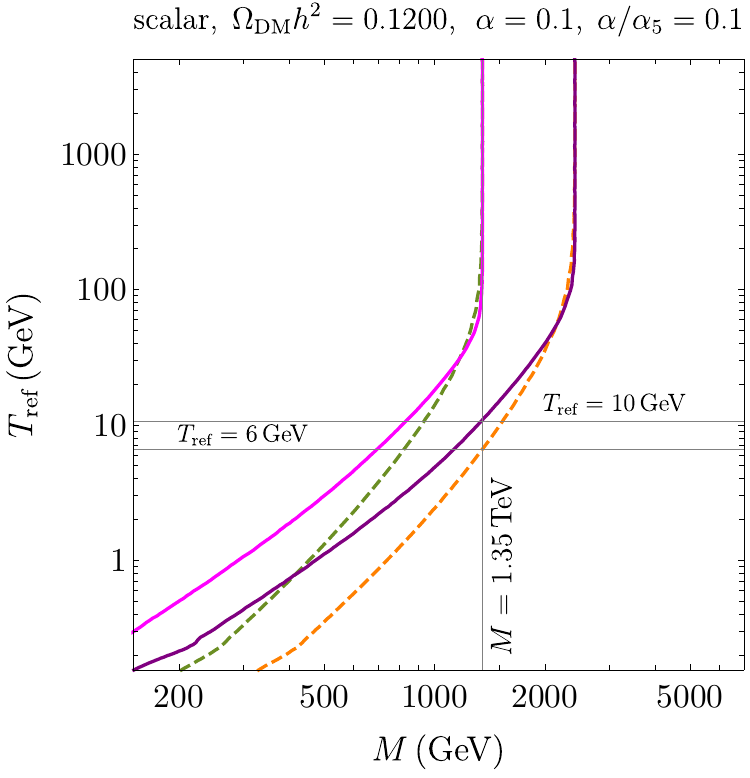}
    \caption{DM energy density contours for $\Omega_{\textrm{DM}}h^2 = 0.012$ in the plane $(M,T_{\textrm{ref}})$ for the vector (left panel) and scalar mediator model (right panels). Here $\alpha=0.1$, $\alpha_5/\alpha=0.1$ and we show two option for the modified cosmologies ($n=2,4$).   }
    \label{fig:M_versus_Tr}
\end{figure}

We take a step forward and now look at the contours that accounts for the Planck measurement of the DM energy density. In figure~\ref{fig:M_versus_Tr}, we explore the $(M,\Tref)$ plane for a fixed coupling strength ($\alpha=0.1$) and for two expansion histories ($n=2,4$).  The DM mass that reproduces the experimental energy density within the standard cosmology can be inferred by looking at the vertical asymptotes for high enough $\Tref$. The two sets of curves in each panel correspond to the DM relic density as obtained with free annihilation cross section, i.e.~without non-perturbative effects, and with Sommerfeld and bound states respectively. Bound-state effects are more prominent for the vector mediator case. For the vector model, the DM mass that is compatible with the observed energy density with the standard cosmology is $M=2.85$ TeV ($M=5.90$ TeV)  for free (non-perturbative) annihilation cross section.   As long as we consider $\Tref \lesssim 200$ GeV, a faster universe expansion demands larger cross sections in order to agree with the observed DM energy density and, therefore, smaller DM masses are needed. The effect of an increasingly longer and faster for expansion is quite important and the DM mass is progressively reduced up to about an order of magnitude for $\Tref = 154$ MeV (we take the QCD phase transition as the minimal reference temperature in figure~\ref{fig:M_versus_Tr}). The faster expansion for $n=4$ bend the contours further towards smaller values of the DM mass. 
Moreover,  one can clearly see how the non-perturbative effects move the contours to larger DM masses (free versus SE+BSF curves), whereas the faster expansion pushes towards smaller masses  for $\Tref \lesssim 200$ GeV.  

There are two observations worth making about some degeneracy that is introduced when considering improved interaction rates and modified expansion histories. We refer to the vector mediator model for the specific values of the parameters. Similar statements hold for the scalar mediator model (mass and $\Tref$ benchmarks are indicated in figure~\ref{fig:M_versus_Tr} right). First, the pair $(M \simeq 800 \; \textrm{GeV},\Tref \simeq 1 \; \textrm{GeV})$ is obtained when using the free annihilation cross section and a modified cosmology with $n=2$ \textit{or} with non-perturbative effects and a modified cosmology with $n=4$ (see intersecting dashed-green and solid-purple curves in figure~\ref{fig:M_versus_Tr}). Second, the dark matter mass $M=2.85$ TeV, which gives the observed energy density for the standard cosmology and free cross section, can be as well obtained with the inclusion of non-perturbative effects and modified cosmologies, respectively for $(\Tref \simeq 9 \; \textrm{GeV},n=2)$ and $(\Tref \simeq 16 \; \textrm{GeV},n=4)$.
\begin{figure}[t!]
    \centering
    \includegraphics[width=0.47\textwidth]{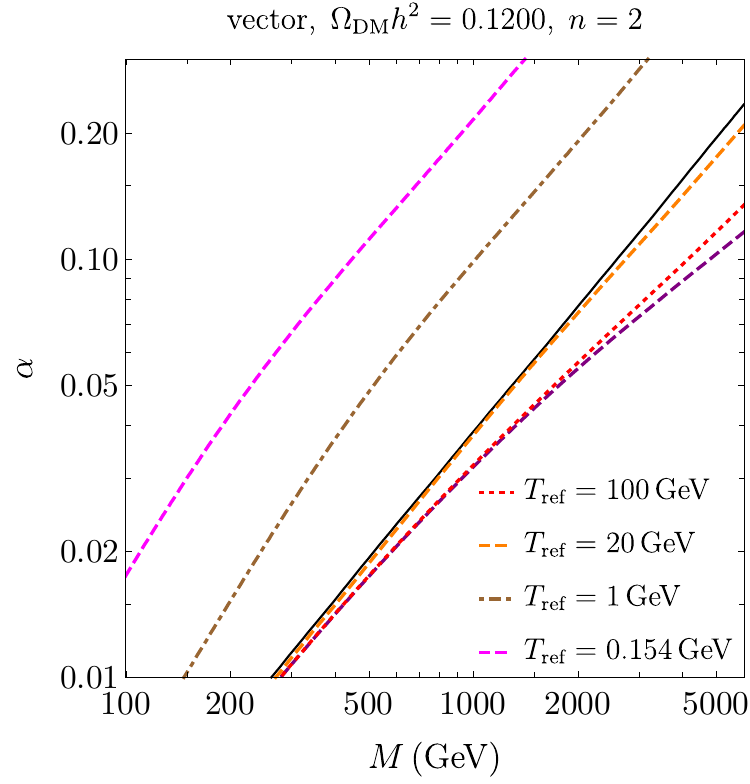}
    \hspace{0.2 cm}
    \includegraphics[width=0.47\textwidth]{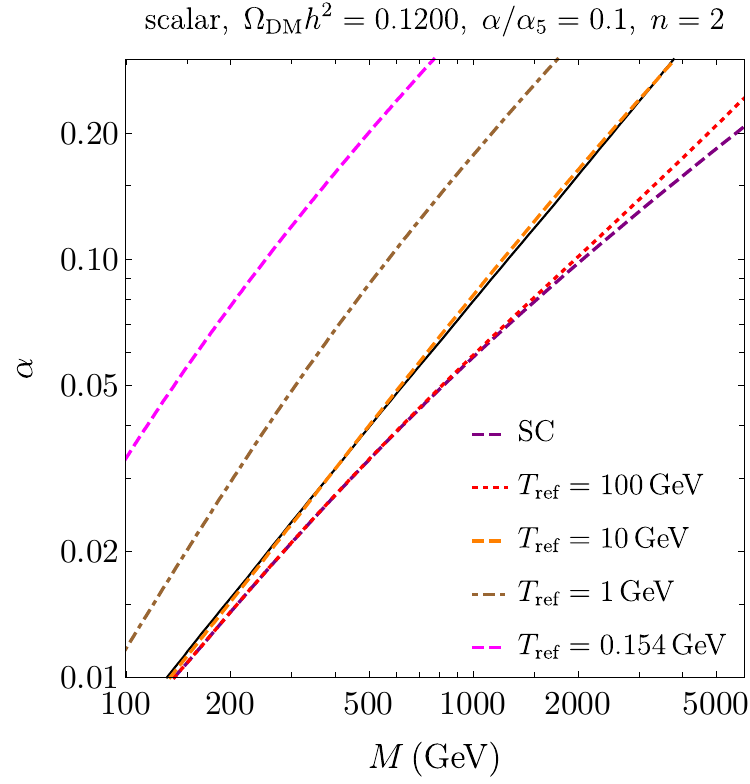}
    \caption{DM energy density contours for $\Omega_{\textrm{DM}}h^2 = 0.1200$ in the parameter space $(M,\alpha)$ for the vector (left panel) and scalar mediator model (right panels). Here we consider different reference temperatures, whereas we fix $n=2$ to identify the kination/quintessence option for a modified expansion history. }
    \label{fig:M_versus_alpha}
\end{figure}

A final look at the the interplay between particle rates and a faster universe expansion is explored in the parameter space $(M,\alpha)$ for different reference temperatures. The results are shown in figure~\ref{fig:M_versus_alpha} for the vector and scalar mediator models.  Here, we restrict to the case $n=2$, which describes the case of kination domination or quintessence. As before, we extract the contours that account for the observed DM energy density, however, we use the annihilation cross section with the inclusion of non-perturbative effects.  The purple-dashed curve is the reference case for the standard cosmology; such curve is also reproduced by modified cosmologies with $\Tref \gtrsim10^3$ GeV and for DM masses $M \lesssim 10$ TeV. Then, for the vector model, the curves for $\Tref=100$ GeV and $\Tref=20$ GeV progressively deviate from the standard cosmology upon increasing the dark matter mass. This is because the freeze-out occurs more towards the epochs when the expansion is different from the standard setting (the freeze-out temperature is proportional to the DM mass). Smaller reference temperatures have a rather large impact and the corresponding curves for $\Tref=1$ GeV and $\Tref=0.154$ GeV shift away from the standard scenario. Here, the observed energy density can be only maintained for large $\alpha$ and small DM masses. For example, for $\alpha=0.1$ in the vector model, modified cosmologies require a DM mass of $M=0.97$ TeV and $M=0.43$ TeV for $\Tref=1$ GeV and $\Tref=0.154$ GeV respectively, instead of the Standard cosmology case $M =4.9$ TeV. 
Finally, we remark a degeneracy of the predicted parameters $(M, \alpha)$ between the reference scenario given by the black-solid-thin line (standard cosmology and \textit{free} annihilation cross section) and modified cosmologies with $\Tref=20$ GeV and $\Tref=10$ GeV for the vector and scalar model respectively, see dashed-orange lines overlapping with the black-solid-thin lines (for the vector case the orange-dashed line gradually detach at large couplings because of more relevant non-perturbative effects with respect to the scalar model). 
\subsection{Freeze-in}
\label{sec:num_FI}
In this section, we present the numerical results for the DM energy density that are extracted with the Born rate with and without thermal masses, cfr.~eqs.~\eqref{fullborneta} and \eqref{vacuum_born}, and its interplay with a modified expansion history of the universe. The rate equation for the DM abundance has been given in eq.~\eqref{number_density_Born}. We further introduce (i) the yield variable $Y_X=n_X/s$; (ii) the actual production rate, that has indeed the dimension of an energy and comprises the Yukawa coupling $y$, together with its thermal average as follows \cite{Asaka:2006rw,Laine:2016hma}
\begin{equation}
    \Gamma(k) = |y|^2\frac{{\rm{Im}}\Pi_{\textrm{R}}(k)}{k_0} \, , \quad \langle \Gamma \rangle \equiv \int \frac{d^3 k}{(2 \pi)^3} \Gamma(k) \,  n_{\textrm{F}}(k_0)  \, .
\end{equation}
Both $\Gamma(k)$ and $\langle \Gamma \rangle$ inherit the temperature dependence of ${\rm{Im}\Pi_R}$ as shown in figure~\ref{fig:IMR_rate_l_and_q}. 
Our integration variable is $x\equiv\ln(T_\mathrm{max}/T)$, with $T_\mathrm{max}$ the maximal temperature, and we start the evolution with vanishing DM abundance $Y_X(x=0)=0$.  We present the numerical results for the top-quark and lepton options. As for the SM couplings, we take them running at one loop (see \cite{Biondini:2020ric} for the renormalisation group equations).

\begin{figure}[t!]
    \centering
       \includegraphics[width=0.47\textwidth]{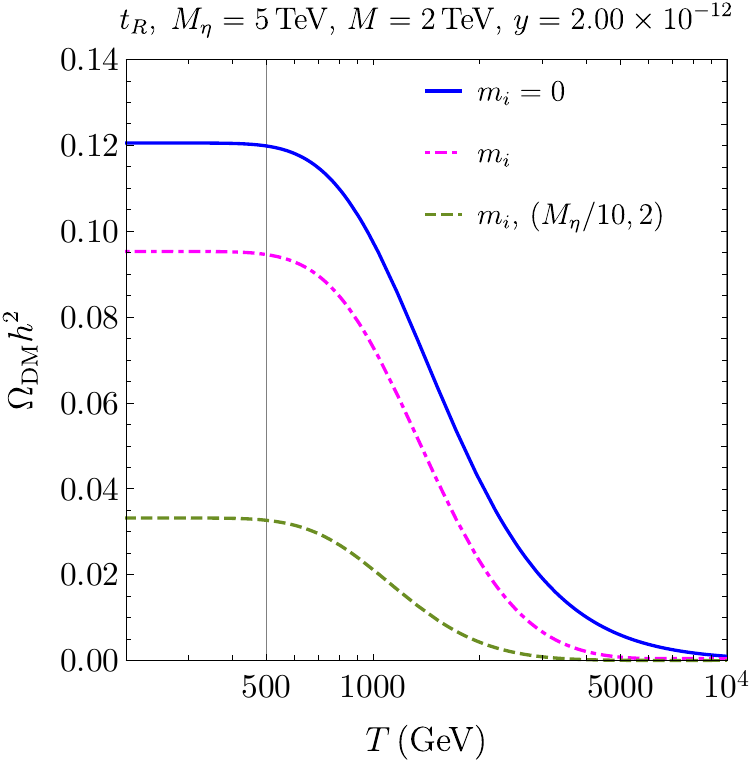}
    \hspace{0.2 cm}
     \includegraphics[width=0.47\textwidth]{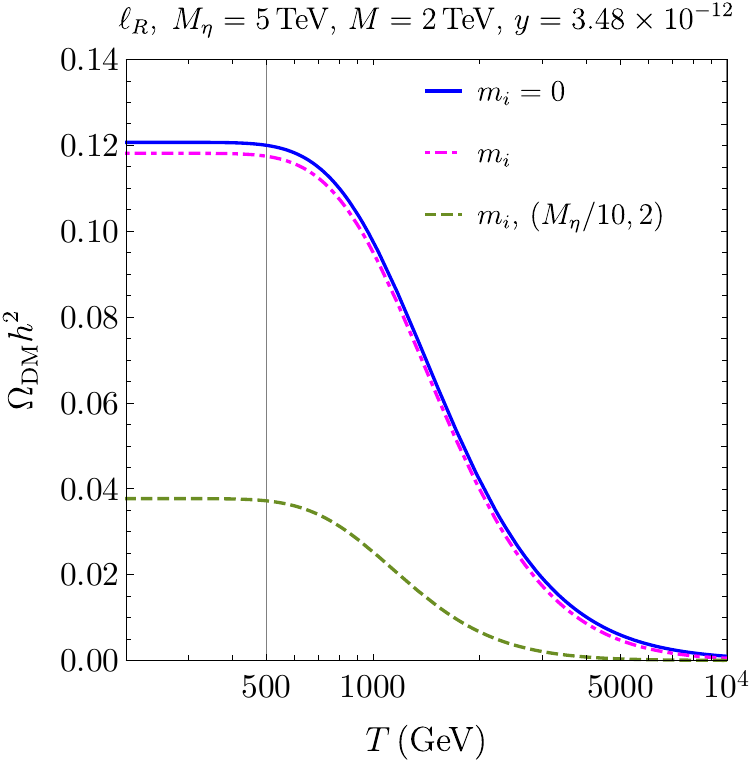}
     \caption{DM energy density as function of the temperature. Thermal and in-vacuum masses are included in the production rate. The parameters of the modified cosmological history are $\Tref=M_\eta/10$ and $n=2$. The mediator-Higgs coupling is $\lambda_3=0$. }
    \label{fig:FI_omega_versus_T_lq}
\end{figure}

In figure~\ref{fig:FI_omega_versus_T_lq} the DM energy density is given as  function of the temperature. One may see how the DM abundance grows from a vanishing initial value and adjusts to a constant, i.e. freezes in, for $T \lesssim M_\eta/10$ as indicated with the gray vertical line. As a reference to single out the impact of a faster universe expansion, the energy density is given for the standard cosmological scenario without and with the inclusion of thermal masses, respectively solid-blue and dot-dashed magenta lines. The Yukawa coupling $y$ is tuned to reproduce the observed $\Omega_{\textrm{DM}}h^2$ for the in-vacuum mass case and with a standard cosmological history. Thermal-mass effects change the final frozen-in density.  The suppression of the production rate with thermal masses (dot-dashed magenta) is more important for the top-quark option, where a correction of about $25\%$ is found, whereas in the lepton case corrections are about few-per cents for our choice of the parameters. When considering the effect of a faster expansion rate with $\Tref=M_\eta/10$ and $n=2$ (dashed-green lines) \textit{and} thermal masses, the predicted energy density drops quite visibly and irrespective of the SM fermion-DM interaction.  It is worth highlighting that the specific choice of the reference temperature $\Tref=M_\eta/10$ makes the faster expansion relevant for the entire duration of the DM production. An important comment is in order. As noticed earlier in ref.~\cite{DEramo:2017ecx}, and at variance with the freeze-out scenario, a faster expansion rate induces a smaller DM energy density as a result of a less effective production rate over the thermal history. Thermal masses makes this feature even more prominent.

\begin{figure}[t!]
    \centering
       \includegraphics[width=0.47\textwidth]{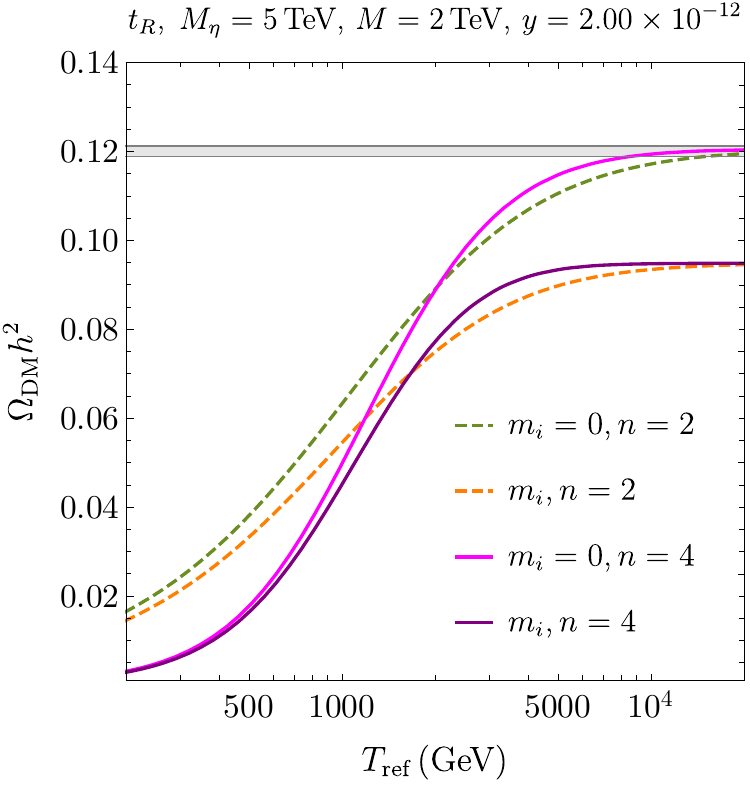}
    \hspace{0.2 cm}
     \includegraphics[width=0.47\textwidth]{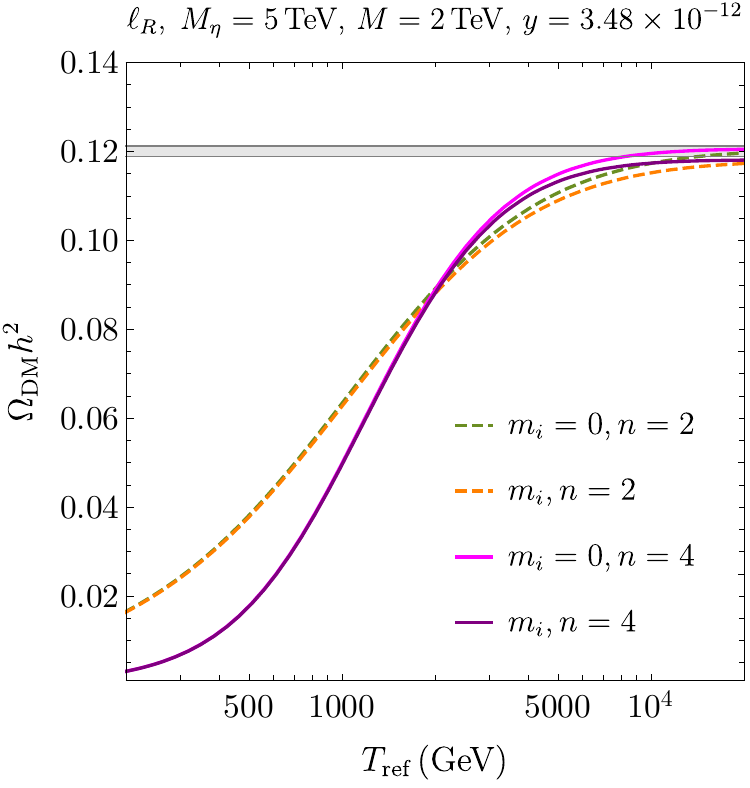}
     \caption{Dark matter energy density as function of $\Tref$, left panel for the top-quark scenario and right-panel for lepto-philic DM. Two cases of modified cosmologies are shown.}
    \label{fig:FI_omega_versus_Tref_lq}
\end{figure}

Finally, in figure~\ref{fig:FI_omega_versus_Tref_lq}, the DM energy density is given as a function of the reference temperature for two values of the modified-cosmology parameter $n=2,4$. Dashed-green and dashed-orange lines stand for $n=2$, whereas solid-magenta and solid-purple curves for $n=4$. The observed DM energy density is attained with vacuum masses for $\Tref \gg M_\eta,M$. Indeed, the values of the Yukawa couplings have been set to obtain $\Omega_{\textrm{DM}}h^2=0.1200$ with in-vacuum production rates and a standard cosmology, see solid-blue lines in figure~\ref{fig:FI_omega_versus_T_lq}.  By decreasing the reference temperature, the effect of a faster expansion becomes more important and reduces the DM energy density of about one order of magnitude for the smallest temperature that we consider, $\Tref=200$ GeV, and for $n=4$.  Moreover, one may see a rather different role of the thermal masses for the two DM-SM fermion interactions. Comparing the left and right panels of figure~\ref{fig:FI_omega_versus_Tref_lq}, one may single out which effect is dominant for the colored or purely-weak interacting mediator. First, the suppression of the production rate from thermal masses is more important for the interaction of the DM with a top quark, as one can notice by looking at the relative separation of the green-dashed and orange-dashed (solid-magenta and solid-purple) lines for $n=2$ ($n=4$). Conversely, the curves are much closer in the case of a lepto-philic DM particle. Second, going towards smaller $\Tref$, there is a progressive approach of the  DM energy density as obtained with in-vacuum or thermal masses. This feature is more visible in the top-quark scenario, and it originates from the shape of the thermal rates ${\rm{Im}}\Pi_R$, see figure~\ref{fig:IMR_rate_l_and_q}). More specifically, in the whole temperature window, that includes the region close to the peak of the particle production, ${\rm{Im}}\Pi_R$ and ${\rm{Im}}\Pi_{R,m_i=0}$ are more far apart for the top-philic case than the corresponding rates for the model with lepton interactions.

\section{Conclusions}
\label{sec:conclusions}
In view of recent advancements of particle interaction rates in the early universe, an update of the cosmologically viable parameter space for various dark matter models is underway. In most cases, this is done by assuming the rather conservative radiation-dominated scenario, which is extrapolated at temperatures much higher than the Big Bang Nucleosynthesis. 
In this paper we have considered the interplay between improved interaction rates, which are the particle-physics input for extracting the DM energy density, and modified cosmological histories. 

For realistic and next-to-minimal DM models, various phenomena can play a role and be relevant when computing the corresponding particle interaction rates. We considered non-perturbative effects for DM thermal freeze-out and the role of thermal masses for freeze-in produced dark matter as triggered by $1 \to 2$ decays. The ameliorated rates can give  corrections as large as one order of magnitude to the predicted DM energy density. We have interfaced improved particle rates with modified cosmological histories in the extraction of the DM energy density. More specifically, we focused on a family of cosmological histories that feature a faster expansion, and that make  particle interactions in the early universe less efficient. 

For dark matter freeze-out we have included near-threshold effects on the annihilations of non-relativistic pairs.  
Here, Sommerfeld factors, bound-state formation and their decays boost DM annihilation and reduce the relic density for a given choice of the model parameters. We have considered fermionic dark matter with a vector or a scalar force mediator and summarized, in the framework of potential non-relativistic effective field theories, the cross sections and widths that enter an effective Boltzmann equation. On the one hand, non-perturbative effects tend to decrease the DM abundance. On the other hand, a faster expansion rate makes the annihilation process less efficient and triggers the opposite trend, namely a larger DM abundance at the freeze-out. We have assessed the interplay between enhanced cross sections and a faster universe expansion through complementary visualizations of the model parameter space. We found that modified cosmologies may open DM mass windows that are compatible with the observed DM energy density. Moreover, there is some degeneracy when extracting the DM energy density with the following two options: (i) standard particle rates and cosmology and (ii) improved annihilation cross sections and modified cosmological histories, see figure~\ref{fig:M_versus_Tr} and \ref{fig:M_versus_alpha}. The latter observation may turn out useful when introducing experimental constraints for the parameter space that is compatible with the observed energy density, that may also apply to modified cosmological histories for specific choice of the reference temperature.  Improvements to the present treatment are in order, such as the inclusion of bound-state effects beyond the no-transition limit and a larger number of excited bound states.

throughout

As for the freeze-in scenario, we have picked one of the various effects that play a role in the ultra-relativistic and relativistic regime, namely thermal masses. At high temperatures, higher than any in-vacuum mass scales, thermal masses modify the two-body phase space and the temperature dependence of the decay processes that produce DM particles. We have inspected the effects of thermal masses for $t$-channel mediators models, that encompass a rich phenomenology even in the case of a feeble interaction between the DM and the visible sector. We found a qualitative difference between a top-philic and lepto-philic DM candidate, which was not noticed in former studies, on the thermal production rate. Owing to rather different Yukawa couplings between SM fermions and the Higgs boson, thermal masses suppress the production rate in the top-quark case, whereas it can be enhanced for DM interacting with leptons. As for the comparison with modified cosmologies, we restrict to choices of the couplings that induce a suppressed production rate in both models. For freeze-in dark matter, a faster expansion history induces an opposite effect with respect to freeze-out. Since the DM abundance builds up over all the thermal history, that include temperatures larger than the DM mass and accompanying state of the dark sector, a faster expansion inhibits the dark matter production and a smaller abundance is found. For the parameters choice of this work, thermal masses add up to a faster expansion in reducing the DM population. For a more rigorous assessment of the interplay with modified cosmologies within this class of models, we remark that the complete set of high-temperature thermal effects, namely multiple-soft scattering  $2 \to 2$ scatterings, should be included, especially when considering smaller mass splittings $\Delta M$.

\section*{Acknowledgment}
This work is supported by the Swiss National Science Foundation (SNSF) under the Ambizione grant PZ00P2\_185783.

\bibliographystyle{hieeetr}
\bibliography{DMcsw}

\begin{thebibliography}{100}

\bibitem{Planck:2018nkj}
N.~Aghanim {\em et~al.}, ``{Planck 2018 results. I. Overview and the
  cosmological legacy of Planck},'' {\em Astron. Astrophys.}, vol.~641, p.~A1,
  2020, 1807.06205.

\bibitem{Bertone:2004pz}
G.~Bertone, D.~Hooper, and J.~Silk, ``{Particle dark matter: Evidence,
  candidates and constraints},'' {\em Phys. Rept.}, vol.~405, pp.~279--390,
  2005, hep-ph/0404175.

\bibitem{Feng:2010gw}
J.~L. Feng, ``{Dark Matter Candidates from Particle Physics and Methods of
  Detection},'' {\em Ann. Rev. Astron. Astrophys.}, vol.~48, pp.~495--545,
  2010, 1003.0904.

\bibitem{Lee:1977ua}
B.~W. Lee and S.~Weinberg, ``{Cosmological Lower Bound on Heavy Neutrino
  Masses},'' {\em Phys. Rev. Lett.}, vol.~39, pp.~165--168, 1977.

\bibitem{Griest:1990kh}
K.~Griest and D.~Seckel, ``{Three exceptions in the calculation of relic
  abundances},'' {\em Phys. Rev.}, vol.~D43, pp.~3191--3203, 1991.

\bibitem{Gondolo:1990dk}
P.~Gondolo and G.~Gelmini, ``{Cosmic abundances of stable particles: Improved
  analysis},'' {\em Nucl. Phys.}, vol.~B360, pp.~145--179, 1991.

\bibitem{Moroi:1993mb}
T.~Moroi, H.~Murayama, and M.~Yamaguchi, ``{Cosmological constraints on the
  light stable gravitino},'' {\em Phys. Lett. B}, vol.~303, pp.~289--294, 1993.

\bibitem{McDonald:2001vt}
J.~McDonald, ``{Thermally generated gauge singlet scalars as selfinteracting
  dark matter},'' {\em Phys. Rev. Lett.}, vol.~88, p.~091304, 2002,
  hep-ph/0106249.

\bibitem{Hall:2009bx}
L.~J. Hall, K.~Jedamzik, J.~March-Russell, and S.~M. West, ``{Freeze-In
  Production of FIMP Dark Matter},'' {\em JHEP}, vol.~03, p.~080, 2010,
  0911.1120.

\bibitem{Sommerfeld}
A.~Sommerfeld, ``{\"Uber die Beugung und Bremsung der Elektronen},'' {\em Ann.
  Phys.(1931)}, vol.~403, 1931.

\bibitem{Hisano:2004ds}
J.~Hisano, S.~Matsumoto, M.~M. Nojiri, and O.~Saito, ``{Non-perturbative effect
  on dark matter annihilation and gamma ray signature from galactic center},''
  {\em Phys. Rev. D}, vol.~71, p.~063528, 2005, hep-ph/0412403.

\bibitem{Detmold:2014qqa}
W.~Detmold, M.~McCullough, and A.~Pochinsky, ``{Dark Nuclei I: Cosmology and
  Indirect Detection},'' {\em Phys. Rev. D}, vol.~90, no.~11, p.~115013, 2014,
  1406.2276.

\bibitem{vonHarling:2014kha}
B.~von Harling and K.~Petraki, ``{Bound-state formation for thermal relic dark
  matter and unitarity},'' {\em JCAP}, vol.~1412, p.~033, 2014, 1407.7874.

\bibitem{Kim:2016kxt}
S.~Kim and M.~Laine, ``{On thermal corrections to near-threshold
  annihilation},'' {\em JCAP}, vol.~1701, p.~013, 2017, 1609.00474.

\bibitem{Biondini:2018pwp}
S.~Biondini and M.~Laine, ``{Thermal dark matter co-annihilating with a
  strongly interacting scalar},'' {\em JHEP}, vol.~04, p.~072, 2018,
  1801.05821.

\bibitem{Binder:2019erp}
T.~Binder, K.~Mukaida, and K.~Petraki, ``{Rapid bound-state formation of Dark
  Matter in the Early Universe},'' {\em Phys. Rev. Lett.}, vol.~124, no.~16,
  p.~161102, 2020, 1910.11288.

\bibitem{Laha:2015yoa}
R.~Laha, ``{Directional detection of dark matter in universal bound states},''
  {\em Phys. Rev. D}, vol.~92, p.~083509, 2015, 1505.02772.

\bibitem{Asadi:2016ybp}
P.~Asadi, M.~Baumgart, P.~J. Fitzpatrick, E.~Krupczak, and T.~R. Slatyer,
  ``{Capture and Decay of Electroweak WIMPonium},'' {\em JCAP}, vol.~02,
  p.~005, 2017, 1610.07617.

\bibitem{Garny:2018ali}
M.~Garny and J.~Heisig, ``{Interplay of super-WIMP and freeze-in production of
  dark matter},'' {\em Phys. Rev.}, vol.~D98, no.~9, p.~095031, 2018,
  1809.10135.

\bibitem{Biondini:2018ovz}
S.~Biondini and S.~Vogl, ``{Coloured coannihilations: Dark matter phenomenology
  meets non-relativistic EFTs},'' {\em JHEP}, vol.~02, p.~016, 2019,
  1811.02581.

\bibitem{Biondini:2019int}
S.~Biondini and S.~Vogl, ``{Scalar dark matter coannihilating with a coloured
  fermion},'' {\em JHEP}, vol.~11, p.~147, 2019, 1907.05766.

\bibitem{Garny:2021qsr}
M.~Garny and J.~Heisig, ``{Bound-state effects on dark matter coannihilation:
  Pushing the boundaries of conversion-driven freeze-out},'' {\em Phys. Rev.
  D}, vol.~105, no.~5, p.~055004, 2022, 2112.01499.

\bibitem{Bottaro:2021srh}
S.~Bottaro, A.~Strumia, and N.~Vignaroli, ``{Minimal Dark Matter bound states
  at future colliders},'' {\em JHEP}, vol.~06, p.~143, 2021, 2103.12766.

\bibitem{Becker:2022iso}
M.~Becker, E.~Copello, J.~Harz, K.~A. Mohan, and D.~Sengupta, ``{Impact of
  Sommerfeld effect and bound state formation in simplified t-channel dark
  matter models},'' {\em JHEP}, vol.~08, p.~145, 2022, 2203.04326.

\bibitem{Biondini:2023ksj}
S.~Biondini, J.~Bollig, and S.~Vogl, ``{Indirect detection of dark matter with
  (pseudo)-scalar interactions},'' 8 2023, 2308.14594.

\bibitem{Belanger:2018ccd}
G.~B\'elanger, F.~Boudjema, A.~Goudelis, A.~Pukhov, and B.~Zaldivar,
  ``{micrOMEGAs5.0 : Freeze-in},'' {\em Comput. Phys. Commun.}, vol.~231,
  pp.~173--186, 2018, 1801.03509.

\bibitem{Lebedev:2019ton}
O.~Lebedev and T.~Toma, ``{Relativistic Freeze-in},'' {\em Phys. Lett. B},
  vol.~798, p.~134961, 2019, 1908.05491.

\bibitem{Bandyopadhyay:2020ufc}
P.~Bandyopadhyay, M.~Mitra, and A.~Roy, ``{Relativistic Freeze-in with Scalar
  Dark Matter in a Gauged $B-L$ Model and Electroweak Symmetry Breaking},'' 12
  2020, 2012.07142.

\bibitem{Anisimov:2010gy}
A.~Anisimov, D.~Besak, and D.~Bodeker, ``{Thermal production of relativistic
  Majorana neutrinos: Strong enhancement by multiple soft scattering},'' {\em
  JCAP}, vol.~1103, p.~042, 2011, 1012.3784.

\bibitem{Besak:2012qm}
D.~Besak and D.~Bodeker, ``{Thermal production of ultrarelativistic
  right-handed neutrinos: Complete leading-order results},'' {\em JCAP},
  vol.~1203, p.~029, 2012, 1202.1288.

\bibitem{Ghisoiu:2014mha}
I.~Ghisoiu and M.~Laine, ``{Interpolation of hard and soft dilepton rates},''
  {\em JHEP}, vol.~10, p.~083, 2014, 1407.7955.

\bibitem{Ghiglieri:2016xye}
J.~Ghiglieri and M.~Laine, ``{Neutrino dynamics below the electroweak
  crossover},'' {\em JCAP}, vol.~1607, no.~07, p.~015, 2016, 1605.07720.

\bibitem{Tenkanen:2016jic}
T.~Tenkanen and V.~Vaskonen, ``{Reheating the Standard Model from a hidden
  sector},'' {\em Phys. Rev. D}, vol.~94, no.~8, p.~083516, 2016, 1606.00192.

\bibitem{Hardy:2018bph}
E.~Hardy, ``{Higgs portal dark matter in non-standard cosmological
  histories},'' {\em JHEP}, vol.~06, p.~043, 2018, 1804.06783.

\bibitem{Bernal:2018kcw}
N.~Bernal, C.~Cosme, T.~Tenkanen, and V.~Vaskonen, ``{Scalar singlet dark
  matter in non-standard cosmologies},'' {\em Eur. Phys. J. C}, vol.~79, no.~1,
  p.~30, 2019, 1806.11122.

\bibitem{Bernal:2018ins}
N.~Bernal, C.~Cosme, and T.~Tenkanen, ``{Phenomenology of Self-Interacting Dark
  Matter in a Matter-Dominated Universe},'' {\em Eur. Phys. J. C}, vol.~79,
  no.~2, p.~99, 2019, 1803.08064.

\bibitem{Arias:2020qty}
P.~Arias, D.~Karamitros, and L.~Roszkowski, ``{Frozen-in fermionic singlet dark
  matter in non-standard cosmology with a decaying fluid},'' {\em JCAP},
  vol.~05, p.~041, 2021, 2012.07202.

\bibitem{Chanda:2019xyl}
P.~Chanda, S.~Hamdan, and J.~Unwin, ``{Reviving $Z$ and Higgs Mediated Dark
  Matter Models in Matter Dominated Freeze-out},'' {\em JCAP}, vol.~01, p.~034,
  2020, 1911.02616.

\bibitem{Bernal:2019mhf}
N.~Bernal, F.~Elahi, C.~Maldonado, and J.~Unwin, ``{Ultraviolet Freeze-in and
  Non-Standard Cosmologies},'' {\em JCAP}, vol.~11, p.~026, 2019, 1909.07992.

\bibitem{Chang:2021ose}
Z.-F. Chang, Z.-X. Chen, J.-S. Xu, and Z.-L. Han, ``{FIMP Dark Matter from
  Leptogenesis in Fast Expanding Universe},'' {\em JCAP}, vol.~06, p.~006,
  2021, 2104.02364.

\bibitem{Barman:2021ifu}
B.~Barman, P.~Ghosh, F.~S. Queiroz, and A.~K. Saha, ``{Scalar multiplet dark
  matter in a fast expanding Universe: Resurrection of the desert region},''
  {\em Phys. Rev. D}, vol.~104, no.~1, p.~015040, 2021, 2101.10175.

\bibitem{Han:2019vxi}
C.~Han, ``{Higgsino Dark Matter in a Non-Standard History of the Universe},''
  {\em Phys. Lett. B}, vol.~798, p.~134997, 2019, 1907.09235.

\bibitem{Salati:2002md}
P.~Salati, ``{Quintessence and the relic density of neutralinos},'' {\em Phys.
  Lett. B}, vol.~571, pp.~121--131, 2003, astro-ph/0207396.

\bibitem{Profumo:2003hq}
S.~Profumo and P.~Ullio, ``{SUSY dark matter and quintessence},'' {\em JCAP},
  vol.~11, p.~006, 2003, hep-ph/0309220.

\bibitem{Drees:2018dsj}
M.~Drees and F.~Hajkarim, ``{Neutralino Dark Matter in Scenarios with Early
  Matter Domination},'' {\em JHEP}, vol.~12, p.~042, 2018, 1808.05706.

\bibitem{Sujuan:2023lne}
Q.~Sujuan and H.~Iminniyaz, ``{Relic density of asymmetric dark matter and
  constraints on the parameter spaces},'' 10 2023, 2310.03711.

\bibitem{Ghosh:2023tyz}
D.~K. Ghosh, A.~Ghoshal, and S.~Jeesun, ``{Axion-like particle (ALP) portal
  freeze-in dark matter confronting ALP search experiments},'' 5 2023,
  2305.09188.

\bibitem{Binder:2018znk}
T.~Binder, L.~Covi, and K.~Mukaida, ``{Dark Matter Sommerfeld-enhanced
  annihilation and Bound-state decay at finite temperature},'' {\em Phys.
  Rev.}, vol.~D98, no.~11, p.~115023, 2018, 1808.06472.

\bibitem{Laine:2006ns}
M.~Laine, O.~Philipsen, P.~Romatschke, and M.~Tassler, ``{Real-time static
  potential in hot QCD},'' {\em JHEP}, vol.~03, p.~054, 2007, hep-ph/0611300.

\bibitem{Brambilla:2008cx}
N.~Brambilla, J.~Ghiglieri, A.~Vairo, and P.~Petreczky, ``{Static
  quark-antiquark pairs at finite temperature},'' {\em Phys. Rev.}, vol.~D78,
  p.~014017, 2008, 0804.0993.

\bibitem{Caswell:1985ui}
W.~E. Caswell and G.~P. Lepage, ``{Effective Lagrangians for Bound State
  Problems in QED, QCD, and Other Field Theories},'' {\em Phys. Lett.},
  vol.~167B, pp.~437--442, 1986.

\bibitem{Bodwin:1994jh}
G.~T. Bodwin, E.~Braaten, and G.~P. Lepage, ``{Rigorous QCD analysis of
  inclusive annihilation and production of heavy quarkonium},'' {\em Phys.
  Rev.}, vol.~D51, pp.~1125--1171, 1995, hep-ph/9407339.
\newblock [Erratum: Phys. Rev.D55,5853(1997)].

\bibitem{Pineda:1997bj}
A.~Pineda and J.~Soto, ``{Effective field theory for ultrasoft momenta in NRQCD
  and NRQED},'' {\em Nucl. Phys. Proc. Suppl.}, vol.~64, pp.~428--432, 1998,
  hep-ph/9707481.

\bibitem{Brambilla:1999xf}
N.~Brambilla, A.~Pineda, J.~Soto, and A.~Vairo, ``{Potential NRQCD: An
  Effective theory for heavy quarkonium},'' {\em Nucl. Phys.}, vol.~B566,
  p.~275, 2000, hep-ph/9907240.

\bibitem{Beneke:2014gja}
M.~Beneke, C.~Hellmann, and P.~Ruiz-Femenia, ``{Non-relativistic pair
  annihilation of nearly mass degenerate neutralinos and charginos III.
  Computation of the Sommerfeld enhancements},'' {\em JHEP}, vol.~05, p.~115,
  2015, 1411.6924.

\bibitem{Binder:2020efn}
T.~Binder, B.~Blobel, J.~Harz, and K.~Mukaida, ``{Dark matter bound-state
  formation at higher order: a non-equilibrium quantum field theory
  approach},'' {\em JHEP}, vol.~09, p.~086, 2020, 2002.07145.

\bibitem{Biondini:2021ycj}
S.~Biondini and V.~Shtabovenko, ``{Bound-state formation, dissociation and
  decays of darkonium with potential non-relativistic Yukawa theory for scalar
  and pseudoscalar mediators},'' {\em JHEP}, vol.~03, p.~172, 2022, 2112.10145.

\bibitem{Biondini:2023zcz}
S.~Biondini, N.~Brambilla, G.~Qerimi, and A.~Vairo, ``{Effective field theories
  for dark matter pairs in the early universe: cross sections and widths},''
  {\em JHEP}, vol.~07, p.~006, 2023, 2304.00113.

\bibitem{Brambilla:2002nu}
N.~Brambilla, D.~Eiras, A.~Pineda, J.~Soto, and A.~Vairo, ``{Inclusive decays
  of heavy quarkonium to light particles},'' {\em Phys. Rev. D}, vol.~67,
  p.~034018, 2003, hep-ph/0208019.

\bibitem{Brambilla:2004jw}
N.~Brambilla, A.~Pineda, J.~Soto, and A.~Vairo, ``{Effective field theories for
  heavy quarkonium},'' {\em Rev. Mod. Phys.}, vol.~77, p.~1423, 2005,
  hep-ph/0410047.

\bibitem{Binder:2021otw}
T.~Binder, K.~Mukaida, B.~Scheihing-Hitschfeld, and X.~Yao, ``{Non-Abelian
  electric field correlator at NLO for dark matter relic abundance and
  quarkonium transport},'' {\em JHEP}, vol.~01, p.~137, 2022, 2107.03945.

\bibitem{Binder:2023ckj}
T.~Binder, M.~Garny, J.~Heisig, S.~Lederer, and K.~Urban, ``{Excited bound
  states and their role in dark matter production},'' 8 2023, 2308.01336.

\bibitem{Feldman:2006wd}
D.~Feldman, B.~Kors, and P.~Nath, ``{Extra-weakly Interacting Dark Matter},''
  {\em Phys. Rev. D}, vol.~75, p.~023503, 2007, hep-ph/0610133.

\bibitem{Fayet:2007ua}
P.~Fayet, ``{U-boson production in e+ e- annihilations, psi and Upsilon decays,
  and Light Dark Matter},'' {\em Phys. Rev. D}, vol.~75, p.~115017, 2007,
  hep-ph/0702176.

\bibitem{Goodsell:2009xc}
M.~Goodsell, J.~Jaeckel, J.~Redondo, and A.~Ringwald, ``{Naturally Light Hidden
  Photons in LARGE Volume String Compactifications},'' {\em JHEP}, vol.~11,
  p.~027, 2009, 0909.0515.

\bibitem{Morrissey:2009ur}
D.~E. Morrissey, D.~Poland, and K.~M. Zurek, ``{Abelian Hidden Sectors at a
  GeV},'' {\em JHEP}, vol.~07, p.~050, 2009, 0904.2567.

\bibitem{Andreas:2011in}
S.~Andreas, M.~D. Goodsell, and A.~Ringwald, ``{Dark matter and dark forces
  from a supersymmetric hidden sector},'' {\em Phys. Rev. D}, vol.~87, no.~2,
  p.~025007, 2013, 1109.2869.

\bibitem{Holdom:1985ag}
B.~Holdom, ``{Two U(1)'s and Epsilon Charge Shifts},'' {\em Phys. Lett. B},
  vol.~166, pp.~196--198, 1986.

\bibitem{Foot:1991kb}
R.~Foot and X.-G. He, ``{Comment on Z Z-prime mixing in extended gauge
  theories},'' {\em Phys. Lett. B}, vol.~267, pp.~509--512, 1991.

\bibitem{Pineda:1997ie}
A.~Pineda and J.~Soto, ``{The Lamb shift in dimensional regularization},'' {\em
  Phys. Lett.}, vol.~B420, pp.~391--396, 1998, hep-ph/9711292.

\bibitem{Vairo:2003gh}
A.~Vairo, ``{A Theoretical review of heavy quarkonium inclusive decays},'' {\em
  Mod. Phys. Lett.}, vol.~A19, pp.~253--269, 2004, hep-ph/0311303.

\bibitem{Pospelov:2007mp}
M.~Pospelov, A.~Ritz, and M.~B. Voloshin, ``{Secluded WIMP Dark Matter},'' {\em
  Phys. Lett. B}, vol.~662, pp.~53--61, 2008, 0711.4866.

\bibitem{Kaplinghat:2013yxa}
M.~Kaplinghat, S.~Tulin, and H.-B. Yu, ``{Direct Detection Portals for
  Self-interacting Dark Matter},'' {\em Phys. Rev. D}, vol.~89, no.~3,
  p.~035009, 2014, 1310.7945.

\bibitem{Wise:2014jva}
M.~B. Wise and Y.~Zhang, ``{Stable Bound States of Asymmetric Dark Matter},''
  {\em Phys. Rev. D}, vol.~90, no.~5, p.~055030, 2014, 1407.4121.
\newblock [Erratum: Phys.Rev.D 91, 039907 (2015)].

\bibitem{Oncala:2018bvl}
R.~Oncala and K.~Petraki, ``{Dark matter bound states via emission of scalar
  mediators},'' {\em JHEP}, vol.~01, p.~070, 2019, 1808.04854.

\bibitem{Kahlhoefer:2017umn}
F.~Kahlhoefer, K.~Schmidt-Hoberg, and S.~Wild, ``{Dark matter self-interactions
  from a general spin-0 mediator},'' {\em JCAP}, vol.~08, p.~003, 2017,
  1704.02149.

\bibitem{Arcadi:2019lka}
G.~Arcadi, A.~Djouadi, and M.~Raidal, ``{Dark Matter through the Higgs
  portal},'' {\em Phys. Rept.}, vol.~842, pp.~1--180, 2020, 1903.03616.

\bibitem{Biondini:2021ccr}
S.~Biondini and V.~Shtabovenko, ``{Non-relativistic and potential
  non-relativistic effective field theories for scalar mediators},'' {\em
  JHEP}, vol.~08, p.~114, 2021, 2106.06472.

\bibitem{Bernal:2017kxu}
N.~Bernal, M.~Heikinheimo, T.~Tenkanen, K.~Tuominen, and V.~Vaskonen, ``{The
  Dawn of FIMP Dark Matter: A Review of Models and Constraints},'' {\em Int. J.
  Mod. Phys. A}, vol.~32, no.~27, p.~1730023, 2017, 1706.07442.

\bibitem{Baker:2016xzo}
M.~J. Baker and J.~Kopp, ``{Dark Matter Decay between Phase Transitions at the
  Weak Scale},'' {\em Phys. Rev. Lett.}, vol.~119, no.~6, p.~061801, 2017,
  1608.07578.

\bibitem{Baker:2017zwx}
M.~J. Baker, M.~Breitbach, J.~Kopp, and L.~Mittnacht, ``{Dynamic Freeze-In:
  Impact of Thermal Masses and Cosmological Phase Transitions on Dark Matter
  Production},'' {\em JHEP}, vol.~03, p.~114, 2018, 1712.03962.

\bibitem{Dvorkin:2019zdi}
C.~Dvorkin, T.~Lin, and K.~Schutz, ``{Making dark matter out of light:
  freeze-in from plasma effects},'' {\em Phys. Rev.}, vol.~D99, no.~11,
  p.~115009, 2019, 1902.08623.

\bibitem{Darme:2019wpd}
L.~Darm\'e, A.~Hryczuk, D.~Karamitros, and L.~Roszkowski, ``{Forbidden
  frozen-in dark matter},'' {\em JHEP}, vol.~11, p.~159, 2019, 1908.05685.

\bibitem{Chakrabarty:2022bcn}
N.~Chakrabarty, P.~Konar, R.~Roshan~and, and S.~Show, ``{Thermally corrected
  masses and freeze-in dark matter: A case study},'' {\em Phys. Rev. D},
  vol.~107, no.~3, p.~035021, 2023, 2206.02233.

\bibitem{Bringmann:2021sth}
T.~Bringmann, S.~Heeba, F.~Kahlhoefer, and K.~Vangsnes, ``{Freezing-in a hot
  bath: resonances, medium effects and phase transitions},'' {\em JHEP},
  vol.~02, p.~110, 2022, 2111.14871.

\bibitem{Garny:2015wea}
M.~Garny, A.~Ibarra, and S.~Vogl, ``{Signatures of Majorana dark matter with
  t-channel mediators},'' {\em Int. J. Mod. Phys.}, vol.~D24, no.~07,
  p.~1530019, 2015, 1503.01500.

\bibitem{DeSimone:2016fbz}
A.~De~Simone and T.~Jacques, ``{Simplified models vs. effective field theory
  approaches in dark matter searches},'' {\em Eur. Phys. J. C}, vol.~76, no.~7,
  p.~367, 2016, 1603.08002.

\bibitem{Arina:2020udz}
C.~Arina, B.~Fuks, and L.~Mantani, ``{A universal framework for t-channel dark
  matter models},'' {\em Eur. Phys. J. C}, vol.~80, no.~5, p.~409, 2020,
  2001.05024.

\bibitem{Belanger:2018sti}
G.~Bélanger {\em et~al.}, ``{LHC-friendly minimal freeze-in models},'' {\em
  JHEP}, vol.~02, p.~186, 2019, 1811.05478.

\bibitem{Drewes:2015eoa}
M.~Drewes and J.~U. Kang, ``{Sterile neutrino Dark Matter production from
  scalar decay in a thermal bath},'' {\em JHEP}, vol.~05, p.~051, 2016,
  1510.05646.

\bibitem{Biondini:2020ric}
S.~Biondini and J.~Ghiglieri, ``{Freeze-in produced dark matter in the
  ultra-relativistic regime},'' {\em JCAP}, vol.~03, p.~075, 2021, 2012.09083.

\bibitem{Garny:2018icg}
M.~Garny, J.~Heisig, M.~Hufnagel, and B.~Lülf, ``{Top-philic dark matter
  within and beyond the WIMP paradigm},'' {\em Phys. Rev.}, vol.~D97, no.~7,
  p.~075002, 2018, 1802.00814.

\bibitem{Junius:2019dci}
S.~Junius, L.~Lopez-Honorez, and A.~Mariotti, ``{A feeble window on leptophilic
  dark matter},'' {\em JHEP}, vol.~07, p.~136, 2019, 1904.07513.

\bibitem{Bollig:2021psb}
J.~Bollig and S.~Vogl, ``{Impact of bound states on non-thermal dark matter
  production},'' {\em JCAP}, vol.~10, p.~031, 2022, 2112.01491.

\bibitem{Asaka:2006rw}
T.~Asaka, M.~Laine, and M.~Shaposhnikov, ``{On the hadronic contribution to
  sterile neutrino production},'' {\em JHEP}, vol.~06, p.~053, 2006,
  hep-ph/0605209.

\bibitem{Bodeker:2015exa}
D.~B\"odeker, M.~Sangel, and M.~W\"ormann, ``{Equilibration, particle
  production, and self-energy},'' {\em Phys. Rev. D}, vol.~93, no.~4,
  p.~045028, 2016, 1510.06742.

\bibitem{Laine:2016hma}
M.~Laine and A.~Vuorinen, {\em {Basics of Thermal Field Theory}}, vol.~925.
\newblock Springer, 2016, 1701.01554.

\bibitem{Kajantie:1995dw}
K.~Kajantie, M.~Laine, K.~Rummukainen, and M.~E. Shaposhnikov, ``{Generic rules
  for high temperature dimensional reduction and their application to the
  standard model},'' {\em Nucl. Phys. B}, vol.~458, pp.~90--136, 1996,
  hep-ph/9508379.

\bibitem{Giudice:2003jh}
G.~F. Giudice, A.~Notari, M.~Raidal, A.~Riotto, and A.~Strumia, ``{Towards a
  complete theory of thermal leptogenesis in the SM and MSSM},'' {\em Nucl.
  Phys. B}, vol.~685, pp.~89--149, 2004, hep-ph/0310123.

\bibitem{Liu:2021mhn}
J.~Liu, X.-P. Wang, and K.-P. Xie, ``{Searching for lepton portal dark matter
  with colliders and gravitational waves},'' {\em JHEP}, vol.~06, p.~149, 2021,
  2104.06421.

\bibitem{Biondini:2022ggt}
S.~Biondini, P.~Schicho, and T.~V.~I. Tenkanen, ``{Strong electroweak phase
  transition in t-channel simplified dark matter models},'' {\em JCAP},
  vol.~10, p.~044, 2022, 2207.12207.

\bibitem{Allahverdi:2010xz}
R.~Allahverdi, R.~Brandenberger, F.-Y. Cyr-Racine, and A.~Mazumdar,
  ``{Reheating in Inflationary Cosmology: Theory and Applications},'' {\em Ann.
  Rev. Nucl. Part. Sci.}, vol.~60, pp.~27--51, 2010, 1001.2600.

\bibitem{Berlin:2016vnh}
A.~Berlin, D.~Hooper, and G.~Krnjaic, ``{PeV-Scale Dark Matter as a Thermal
  Relic of a Decoupled Sector},'' {\em Phys. Lett. B}, vol.~760, pp.~106--111,
  2016, 1602.08490.

\bibitem{Berlin:2016gtr}
A.~Berlin, D.~Hooper, and G.~Krnjaic, ``{Thermal Dark Matter From A Highly
  Decoupled Sector},'' {\em Phys. Rev. D}, vol.~94, no.~9, p.~095019, 2016,
  1609.02555.

\bibitem{Dine:1995uk}
M.~Dine, L.~Randall, and S.~D. Thomas, ``{Supersymmetry breaking in the early
  universe},'' {\em Phys. Rev. Lett.}, vol.~75, pp.~398--401, 1995,
  hep-ph/9503303.

\bibitem{Caldwell:1997ii}
R.~R. Caldwell, R.~Dave, and P.~J. Steinhardt, ``{Cosmological imprint of an
  energy component with general equation of state},'' {\em Phys. Rev. Lett.},
  vol.~80, pp.~1582--1585, 1998, astro-ph/9708069.

\bibitem{Sahni:1999gb}
V.~Sahni and A.~A. Starobinsky, ``{The Case for a positive cosmological Lambda
  term},'' {\em Int. J. Mod. Phys. D}, vol.~9, pp.~373--444, 2000,
  astro-ph/9904398.

\bibitem{DEramo:2017gpl}
F.~D'Eramo, N.~Fernandez, and S.~Profumo, ``{When the Universe Expands Too
  Fast: Relentless Dark Matter},'' {\em JCAP}, vol.~05, p.~012, 2017,
  1703.04793.

\bibitem{Wetterich:1987fm}
C.~Wetterich, ``{Cosmology and the Fate of Dilatation Symmetry},'' {\em Nucl.
  Phys. B}, vol.~302, pp.~668--696, 1988, 1711.03844.

\bibitem{Dimopoulos:2017zvq}
K.~Dimopoulos and C.~Owen, ``{Quintessential Inflation with
  $\alpha$-attractors},'' {\em JCAP}, vol.~06, p.~027, 2017, 1703.00305.

\bibitem{Kamenshchik:2001cp}
A.~Y. Kamenshchik, U.~Moschella, and V.~Pasquier, ``{An Alternative to
  quintessence},'' {\em Phys. Lett. B}, vol.~511, pp.~265--268, 2001,
  gr-qc/0103004.

\bibitem{Chavanis:2014lra}
P.-H. Chavanis, ``{Cosmology with a stiff matter era},'' {\em Phys. Rev. D},
  vol.~92, no.~10, p.~103004, 2015, 1412.0743.

\bibitem{Khoury:2001wf}
J.~Khoury, B.~A. Ovrut, P.~J. Steinhardt, and N.~Turok, ``{The Ekpyrotic
  universe: Colliding branes and the origin of the hot big bang},'' {\em Phys.
  Rev. D}, vol.~64, p.~123522, 2001, hep-th/0103239.

\bibitem{Davoudiasl:2023khm}
H.~Davoudiasl and M.~Sullivan, ``{Adagio for Thermal Relics},'' 8 2023,
  2308.10928.

\bibitem{Hannestad:2004px}
S.~Hannestad, ``{What is the lowest possible reheating temperature?},'' {\em
  Phys. Rev. D}, vol.~70, p.~043506, 2004, astro-ph/0403291.

\bibitem{HotQCD:2014kol}
A.~Bazavov {\em et~al.}, ``{Equation of state in ( 2+1 )-flavor QCD},'' {\em
  Phys. Rev. D}, vol.~90, p.~094503, 2014, 1407.6387.

\bibitem{Ellis:2015vaa}
J.~Ellis, F.~Luo, and K.~A. Olive, ``{Gluino Coannihilation Revisited},'' {\em
  JHEP}, vol.~09, p.~127, 2015, 1503.07142.

\bibitem{Binder:2021vfo}
T.~Binder, A.~Filimonova, K.~Petraki, and G.~White, ``{Saha equilibrium for
  metastable bound states and dark matter freeze-out},'' {\em Phys. Lett. B},
  vol.~833, p.~137323, 2022, 2112.00042.

\bibitem{Laine:2015kra}
M.~Laine and M.~Meyer, ``{Standard Model thermodynamics across the electroweak
  crossover},'' {\em JCAP}, vol.~07, p.~035, 2015, 1503.04935.

\bibitem{DEramo:2017ecx}
F.~D'Eramo, N.~Fernandez, and S.~Profumo, ``{Dark Matter Freeze-in Production
  in Fast-Expanding Universes},'' {\em JCAP}, vol.~02, p.~046, 2018,
  1712.07453.

\end{thebibliography}


\end{document}